\documentclass[journal=jctcce,manuscript=article]{achemso}

\usepackage[version=3]{mhchem} 

\SectionNumbersOn

\usepackage{amsmath}
\usepackage{amssymb}
\usepackage{color}
\usepackage{CJK}
\usepackage{framed}
\usepackage{ulem}
\usepackage[colorlinks=false,hidelinks]{hyperref}
\usepackage{algorithm}
\usepackage{algpseudocode}
\algdef{SE}[DOWHILE]{Do}{doWhile}{\algorithmicdo}[1]{\algorithmicwhile\ #1}

\usepackage[scaled=0.85]{beramono}
\usepackage[T1]{fontenc}
\usepackage{changepage}
\usepackage{cleveref}
\usepackage{subcaption}
\usepackage{float}
\usepackage{xr} 

\usepackage[disable]{todonotes} 

\author{Nicholas S.M. Herringer}
\affiliation{%
  Department of Chemistry, %
  University of Chicago, %
  Chicago, Illinois 60637%
}

\author{Siva Dasetty}
\affiliation{%
  Pritzker School of Molecular Engineering, %
  University of Chicago, %
  Chicago, Illinois 60637%
}

\author{Diya Gandhi}
\affiliation{%
  Pritzker School of Molecular Engineering, %
  University of Chicago, %
  Chicago, Illinois 60637%
}

\author{Junhee Lee}
\affiliation{%
  Pritzker School of Molecular Engineering, %
  University of Chicago, %
  Chicago, Illinois 60637%
}

\author{Andrew L. Ferguson}
\email{andrewferguson@uchicago.edu}
\affiliation{%
  Pritzker School of Molecular Engineering, %
  University of Chicago, %
  Chicago, Illinois 60637%
}

\title[]{Permutationally Invariant Networks for Enhanced Sampling (PINES): Discovery of Multi-Molecular and Solvent-Inclusive Collective Variables}

\begin{document}


\newpage

\begin{abstract}

\noindent The typically rugged nature of molecular free energy landscapes can frustrate efficient sampling of the thermodynamically relevant phase space due to the presence of high free energy barriers. Enhanced sampling techniques can improve phase space exploration by accelerating sampling along particular collective variables (CVs). A number of techniques exist for data-driven discovery of CVs parameterizing the important large scale motions of the system. A challenge to CV discovery is learning CVs invariant to symmetries of the molecular system, frequently rigid translation, rigid rotation, and permutational relabeling of identical particles. Of these, permutational invariance have proved a persistent challenge in frustrating the the data-driven discovery of multi-molecular CVs in systems of self-assembling particles and solvent-inclusive CVs for solvated systems. In this work, we integrate Permutation Invariant Vector (PIV) featurizations with autoencoding neural networks to learn nonlinear CVs invariant to translation, rotation, and permutation, and perform interleaved rounds of CV discovery and enhanced sampling to iteratively expand sampling of configurational phase space and obtain converged CVs and free energy landscapes. We demonstrate the Permutationally Invariant Network for Enhanced Sampling (PINES) approach in applications to the self-assembly of a 13-atom Argon cluster, association/dissociation of a NaCl ion pair in water, and hydrophobic collapse of a \ce{C_{45}H_{92}} $n$-pentatetracontane polymer chain. We make the approach freely available as a new module within the PLUMED2 enhanced sampling libraries.

\end{abstract}


\newpage

\section{\label{sec:intro}Introduction}

Classical molecular dynamics (MD) simulations present a powerful technique to understand and engineer the molecular-level behaviors of chemical, materials, and biological systems.\cite{AllenTildesley, FrenkelSmit} Rugged free energy landscapes are a generic feature of many molecular systems \cite{dill1997levinthal} wherein high free energy barriers several multiples or more of the thermal energy can trap unbiased simulation trajectories and result in incomplete sampling of the thermally accessible phase space, failure to resolve the relevant metastable states, and unconverged thermodynamic averages \cite{Shirts2022SamplingMethods,Gai2019EnhancedSampling}. Collective variable (CV)-based biasing techniques are a class of non-Boltzmann sampling techniques designed to ameliorate the sampling problem by accelerating barrier crossing along particular CVs \cite{abrams2013enhanced}. A panoply of CV biasing enhanced sampling techniques exist \cite{Ferguson2020MLCVDiscovery,Ferguson2018MESA,Shirts2022SamplingMethods,Gai2019EnhancedSampling}, including, for example, umbrella sampling\cite{Valleau:1977:Umbrella}, adaptive biasing force\cite{Pohorille2001ABF}, and metadynamics\cite{Parrinello2002Meta}. The techniques differ in the numerical details of how the biasing forces are applied, but the success of all of these techniques relies upon the identification of appropriate CVs along which to drive sampling. Robert Zwanzig's observation regarding the projection operator formalism is apposite \cite{zwanzig2001nonequilibrium}: ``Statistical mechanics does not tell us what the relevant variables are. This is our choice. If we choose well, the results may be useful; if we choose badly, the results (while still formally correct) will probably be useless.'' Generally, we wish to select CVs coincident with the large scale and/or slow motions of the system so that we can enhance sampling along the conformational degrees of freedom important to the system dynamics. We also wish to judiciously select a parsimonious number of CVs so as to enable efficient sampling and the opportunity for thermodynamically converged sampling within the CV subspace. CV discovery is fundamentally a problem of dimensionality reduction in the space of the Cartesian coordinates defining the configurational phase space of the system, and these objective functions -- high-variance CVs for large scale motions or maximally autocorrelated CVs for slow motions -- underpin the preponderance of systematic CV discovery techniques using variants of dimensionality reduction algorithms \cite{wang2018nonlinear,Ferguson2018MESA,Ferguson2020MLCVDiscovery}.

Autoencoders -- artificial neural networks (ANNs) comprising an encoding of high-dimensional observations into a low-dimensional latent bottleneck followed by decoding of the latent space back up to the high-dimensional observational space -- have proven particularly well-suited to CV discovery applications \cite{kingma2019introduction,kingma2013auto}. As universal function approximators capable of unsupervised learning and automatic differentiation, these architectures can be trained over unlabeled simulation trajectories to discover nonlinear CVs that are explicit and differentiable functions of the atomic coordinates \cite{Ferguson2020MLCVDiscovery}. These favorable properties mean that it is not necessary to know the metastable states of the system \textit{a priori} (i.e., unlabeled trajectory data is sufficient) and the differentiable CVs can be modularly incorporated into standard enhanced sampling algorithms by propagating biasing forces on the CVs to those on the atoms using the chain rule \cite{Ferguson2018MESA,Ferguson2018CVDiscovery}. In terms of high-variance CV discovery, we previously introduced Molecular Enhanced Sampling with Autoencoders (MESA) as an approach that interleaves successive rounds of CV learning and biased sampling to simultaneously discover CVs coincident with the large-scale dynamical motions of the system and enhance sampling along these coordinates \cite{Ferguson2018MESA,Ferguson2018CVDiscovery}. The process is iterated until we achieve convergence in phase space exploration and a set of CVs that globally parameterize the thermally relevant configurational subspace. Tiwary and co-workers developed reweighted autoencoded variational Bayes for enhanced sampling (RAVE)\cite{Tiwary2018RAVE} as an approach similar to MESA that focusses on the discovery of a single CV that is highly correlated with a physically interpretable order parameter. Belkacemi, Gkeka, Leli\`{e}vre, and Stoltz proposed free energy biasing and iterative learning with autoencoders (FEBILAE) as a technique very closely related to MESA but which reweights the biased simulation data prior to each new round of CV discovery to eliminate the effect of non-Boltzmann sampling in the learned CVs \cite{Stolz2021FEBILAE}. Ensing and co-workers developed the FABULOUS approach that shares similarities to both MESA and RAVE as well as earlier work by Ma and Dinner \cite{hooft2021discovering,ma2005automatic}. This approach essentially employs only the decoding block of an autoencoder to identify CVs capable of accurate reconstruction of atomic coordinates. Luber and co-workers developed deep autoencoder neural network (DAENN)\cite{Luber:2022:DAENN} and a framework called Deep Learning for Collective Variables (DeepCV)\cite{Luber:2022:DeepCV} implementing it to iteratively learn CVs of chemical reactions from limited training configurations covering only the reactant state using eXtended Social PeRmutation InvariaNT (xSPRINT) molecular representation. In terms of slow CV discovery, a number of unsupervised autoencoder-based architectures have also found success, including the time-lagged autoencoders (TAE) of Wehmeyer and No\'{e} \cite{Noe2018TLAE}, variational dynamics encoders (VDE) of Pande and co-workers \cite{hernandez2018variational,sultan2018transferable,wayment2018note}, state predictive information bottleneck (SPIB) of Wang and Tiwary \cite{TiwarySPIB}, and our state-free reversible VAMPNets (SRVs) \cite{chen2019nonlinear,chen2019capabilities,Ferguson:23:GREST} also known as Deep-TICA \cite{bonati2021deep}. In a supervised learning modality, Bolhuis and co-workers employed augmented/extended autoencoders that also take in labeled information on the molecular committor function to learn CVs capable of accurate reconstruction of both the atomic coordinates and progression along a reactive path \cite{Bolhuis:21:JCP}. 

An important consideration in CV discovery is the determination of symmetry-adapted CVs that respect the underlying symmetries of the molecular system \cite{musil2021physics,himanen2020dscribe}. The most commonly encountered symmetries are translation, rotation, and permutation arising from the invariance of a molecular system to rigid translation and/or rotation and the permutational relabeling of identical and indistinguishable particles. Failure to respect these symmetries in the learned CVs is at best data inefficient and at worst can lead to fundamental failure modes associated with the artificial distinction of physically identical system configurations. In general, CVs for enhanced sampling should be invariant to these symmetries, although there are cases where covariance may be desirable \cite{anderson2019cormorant,smidt2021euclidean,batzner20223}. Symmetry adaptation is typically enforced by the choice of featurization of the system and/or within the neural network architecture. The learned CVs are relatively easily made invariant to translation and rotation by representing the system using internal coordinates (i.e., bonds, angles, torsions) that do not depend on the arbitrary location and orientation of the Cartesian coordinate system \cite{musil2021physics,sittel2014principal,zhang2018deep} or by data augmentation \cite{Ferguson2018MESA,shorten2019survey}. 

Permutational invariance has proven more challenging to eliminate, and this has frustrated CV discovery and biased sampling in common and important applications such as self-assembling systems of identical particles or solutes immersed in a bath of identical solvent molecules. Good methods do exist, however, to handle permutational invariance. At the level of the network architecture, Winter, No\'{e}, and Clevert developed a permutation-invariant graph autoencoder (PIGAE) for graph reconstruction by employing an explicit permuter that learns an explicit permutation matrix for each input graph, although the quartic scaling in reordering operations during each training step makes this approach expensive \cite{Noe2021PIVAutoencoder}. Similarly, Huang and co-workers\cite{Xuhui:2023:graphvampnets} developed a GraphVAMPnets based approach that utilizes graph neural networks to respect permutation and rotational symmetries of particles in self-assembling systems by enforcing identical node embeddings for such symmetric particles and learn the slow CVs from the resulting graph embeddings using VAMPnets. Prudente, Acioli, and Soares Neto \cite{prudente1998fitting} and Nguyen and Le \cite{nguyen2012modified} developed special purpose neural networks for the fitting of potential energy surfaces of small molecules in which the first layer of the network was modified to respect permutational invariance, although generalization and scaling to larger systems is challenging. A cheaper and more scalable approach is to represent the system to the network via permutationally-invariant featurizations \cite{musil2021physics}. Provided these featurizations can be expressed as explicit functions of the particle coordinates, the derivatives of the CVs learned by the neural network with respect to the Cartesian coordinates can be constructed from the chain rule via automatic differentiation and/or analytical derivatives. Permutational invariance is typically enforced by summing particle-centric basis functions over all identical particles or a permutationally-invariant ordering. Examples of the former class include permutation-invariant polynomials (PIPs) generated by summing monomial functions of internal coordinates over all like-particle orderings \cite{braams2009permutationally,xie2010permutationally}, atom-centered symmetry functions (ACSF) that achieve permutational invariance by a similar summation procedure \cite{behler2007generalized,behler2011atom}, many-body tensor representations (MBTR) that define scalar geometry functions (e.g., atomic number, distance, angle) over small body-order numbers of atoms and arranges these into permutationally-invariant distribution functions for each class of interaction \cite{huo2022unified}, and smooth overlap of atomic orbitals (SOAP) that generates the partial power spectra of an expansion of spherical harmonic representations of the atomic-centered smoothed density fields summed over all identical atoms \cite{bartok2013representing}. In general, the exponential scaling in the number of permutational orderings associated with the summations for each of these approaches must be controlled by defining a cutoff in the range and/or order of the interactions considered \cite{musil2021physics}. Ordering approaches typically offer a computationally cheaper means to achieve permutational symmetry adaptation \cite{musil2021physics}. Examples of this class include the bag-of-bonds (BoB) featurization based on a concatenation of the ordered (inverse) pairwise distances for each class of bond scaled by the product of the nuclear charges \cite{hansen2015machine} and social permutation invariant (SPRINT) coordinates constructed from the leading eigenvector of a graph adjacency matrix \cite{pietrucci2011graph,Pietrucci2020PIV}. Of particular interest in this work is the Permutation Invariant Vectors (PIV) featurization developed by Pietrucci and co-workers \cite{pipolo2017navigating,Pietrucci2013PIV,Pietrucci2020PIV}. This representation can be conceived of as a concatenation of instantaneous radial distribution functions for classes of atom pairs in the system that, by virtue of distance-based ordering, is invariant to like-particle permutation and, through the use of exclusively pairwise distances, also translation and rotation. Physically motivated, bespoke order parameters engineered to preserve permutational invariance have also been developed, although these can be less generically extensible to arbitrary molecular systems. A trivial strategy to treat the permutational invariance of a solvent bath is to simply ignore the solvent coordinates and assume that the solvent effects are sufficiently strongly imprinted on the solute conformational dynamics that they can be driven implicitly by biasing solute-centric CVs \cite{ferguson2010sys}. Patel, Varilly, Chandler, and Garde developed the indirect umbrella sampling (INDUS) approach to drive water density fluctuations in user-defined control volumes and have used this technique to drive sampling of wetting/dewetting transitions \cite{patel2011quantifying,Patel2021PolymerSolvationINDUS}. Zou \textit{et al.}\ employed CVs based on coordination numbers, Steinhardt bond order parameters, interfacial water density, intermolecular angles, and pair orientational entropy, and combined these within the SPIB framework to drive sampling of crystal nucleation and polymorph formation \cite{Tiwary2022NucleationSPIB}. Rizzi \textit{et al.}\ combined CVs describing solvent coordination number and solvent structure around a binding pocket with the deep linear discriminant analysis (Deep-LDA) approach to incorporate water coordinates in driving sampling of host-guest interactions \cite{Parrinello2021HostGuestWater}.

The primary contribution of this work is to present an approach for the integrated data-driven discovery of translationally, rotationally, and permutationally-invariant CVs in systems containing identical particles and their use within enhanced sampling calculations to accelerate phase space exploration. We term this approach Permutationally Invariant Networks for Enhanced Sampling (PINES). Our approach employs a PIV representation as input for an autoencoder to perform interleaved nonlinear CV discovery and CV biasing enhanced sampling using parallel bias metadynamics \cite{Pfaendtner2015PBMD}. We draw inspiration for the synthesis of permutationally invariant featurizations with ANNs from the deep potential molecular dynamics (DeePMD) approach of Zhang \textit{et al.} that employs an ordered distance featurization passed to an ANN to learn a potential energy function for molecular dynamics \cite{zhang2018deep} and the PIP-NN approach of Jiang and Guo that performs a similar feat using a permutation-invariant polynomial (PIP) featurization \cite{jiang2013permutation}. We eschew the use of physically-motivated, hand-crafted descriptors in favor of a more generic, transferable, and system agnostic featurization strategy, but observe that the incorporation of system-specific knowledge may be straightforwardly achieved by augmenting the descriptor list with any additional features provided they are explicit and differentiable functions of the particle Cartesian coordinates. The PINES paradigm is also modular in the sense that different featurizations may be employed to adapt to different symmetries of the system (e.g., translation, rotation, permutation, inversion, chiral) and different ANNs may be used to perform different learning tasks (e.g., unsupervised CV discovery, supervised learning of potential energy surfaces). We demonstrate PINES in applications to three molecular systems in which proper accounting for particle permutational invariance is of paramount concern: (i) self-assembly of a 13-atom \ce{Ar} cluster, (ii) assembly/disassembly of a \ce{NaCl} ion pair in water, and (iii) hydrophobic collapse of a \ce{C_{45}H_{92}} $n$-pentatetracontane polymer chain. We make the approach freely available as a new user-friendly module at \url{https://github.com/Ferg-Lab/pines.git} that can be patched with PLUMED2 \cite{PLUMED2} enhanced sampling libraries.

The structure of the remainder of this paper is as follows. In Sec.~\ref{sec:methods}, we introduce the PINES method for translationally, rotationally, and permutationally-invariant CV discovery and enhanced sampling using parallel bias metadynamics. This is followed by a description of the PIV featurization, PINES training procedure, and the enhanced sampling molecular dynamics simulations. In Sec.~\ref{sec:results}, we demonstrate applications of PINES in discovering CVs for enhanced sampling to self-assembly of 13-atom \ce{Ar} cluster, association/disassociation of \ce{NaCl} ion pair in water, and hydrophobic collapse of \ce{C_{45}H_{92}} polymer chain, and compare their free energy landscapes to prior work. Finally, in Sec.~\ref{sec:conclusions} we present our conclusions and perspectives for future work.

\section{\label{sec:methods}Methods}

\subsection{Permutationally Invariant Networks for Enhanced Sampling (PINES)}

\subsubsection{Molecular Enhanced Sampling with Autoencoders (MESA)}

\begin{figure}[ht!]
\centering
  \includegraphics[width=0.88\textwidth]{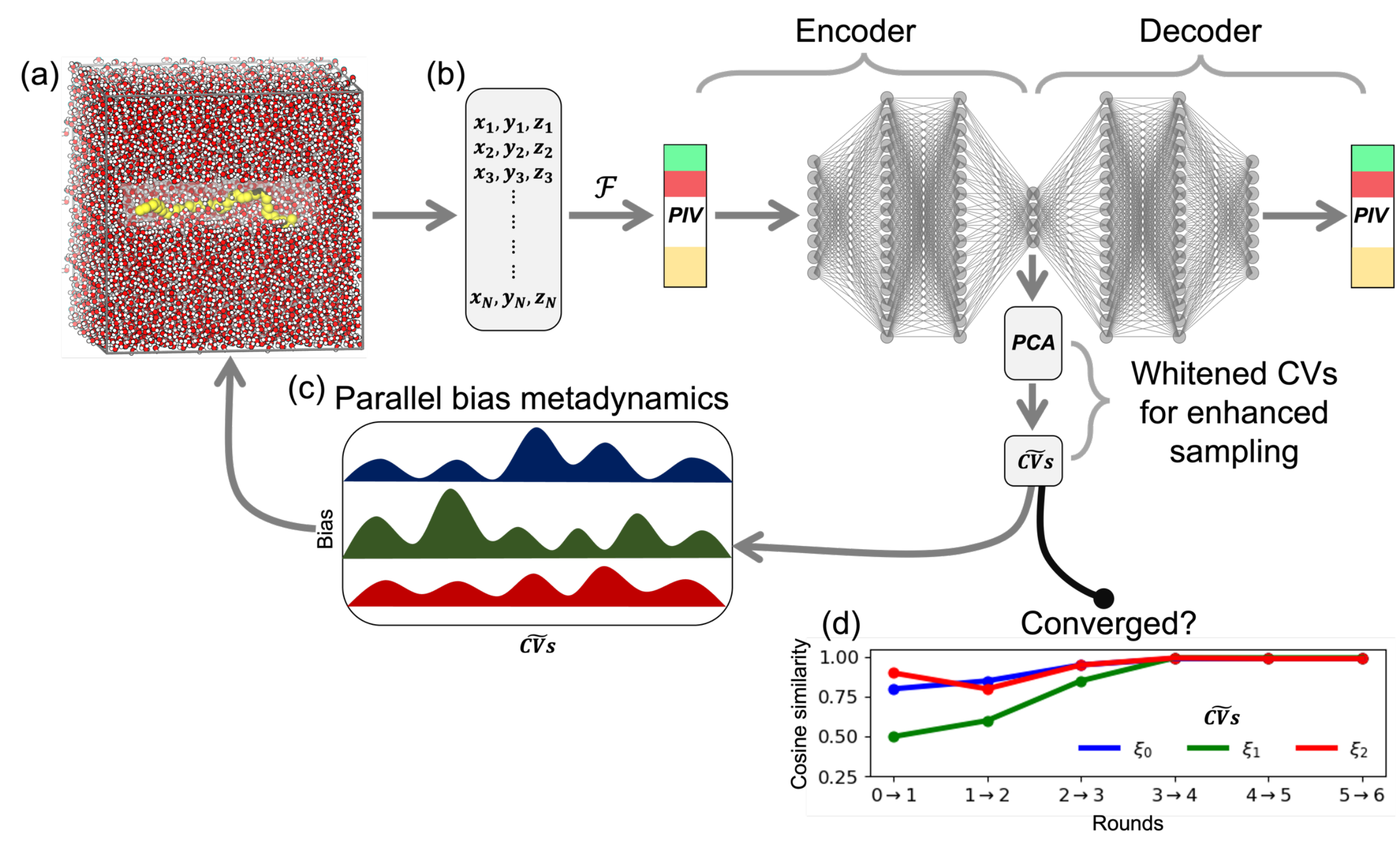}
  \caption{Schematic illustration of the Permutationally Invariant Network for Enhanced Sampling (PINES) approach for data-driven CV discovery and enhanced sampling. (a) Molecular simulation trajectories provide training data for CV discovery. The process is initialized with unbiased or biased simulations to provide an initial first sampling of phase space. Subsequent rounds are trained over biased simulations generated by enhanced sampling in the learned CVs from the previous round. (b) The trajectories are featurized and passed to a maximum mean discrepancy-Wasserstein autoencoder (MMD-WAE) to perform unsupervised nonlinear dimensionality reduction into a low-dimensional latent space spanned by the learned CVs. To stabilize the CV discovery process and help disentangle the latent space, we employ batch normalization between all layers with exception of the final layer. (c) After training the network, we extract the CVs from the bottleneck layer between the encoder and decoder. We find that it is useful to first pass the output of the bottleneck layer through a principal components analysis (PCA) whitening transformation to help align the discovered manifold with the highest variance directions for efficient parallel bias metadynamics (PBMetaD) sampling. The learned CVs are represented as the functional composition $\boldsymbol{\xi} = h \circ f (\boldsymbol{r}) = h(\boldsymbol{z})$, where $\boldsymbol{z} = f (\boldsymbol{r})$ is the result of applying the featurization $f$ -- in this work a PIV featurization -- to the Cartesian coordinates $\boldsymbol{r}$ of the system, and $\boldsymbol{\xi} = h(\boldsymbol{z})$ is the action of the MMD-WAE encoder plus PCA whitening transform upon the input featurization that generates a projection into the learned CVs. The derivative of the CVs with respect to the Cartesian coordinates of atom $i$ is required for propagation of biasing forces in the CVs to biasing forces on the atoms, and comes from simple application of the chain rule to the function composition that is computed in practice by automatic differentiation, $\nabla_{\boldsymbol{r}_{i}} \boldsymbol{\xi} = (\partial h / \partial \boldsymbol{z}) \nabla_{\boldsymbol{r}_{i}} \boldsymbol{z}$. (d) The learned CVs are implemented within PBMetaD calculations to drive sampling in the configurational phase space spanned by these CVs, completing one full iteration of PINES. The biased simulation trajectories from Round $n$ are then passed forward as the training trajectories for CV discovery within Round $(n+1)$ to determine new CVs and conduct new enhanced sampling calculations. The iterative process is terminated when the learned CVs stabilize, observation of no further exploration of phase space, and the reweighted free energy landscape converges.}
  \label{fig:PINESflowchart}
\end{figure}

PINES can be conceived as a permutationally-invariant extension to the Molecular Enhanced Sampling with Autoencoders (MESA) approach that we previously developed for data-driven CV discovery and enhanced sampling \cite{Ferguson2018MESA,Ferguson2018CVDiscovery}. The key innovation of this approach is to use autoencoding ANNs to perform unsupervised, nonlinear dimensionality reduction of high-dimensional molecular simulation trajectories to discover low-dimensional CV parameterizations spanning the high-variance collective motions of the system. By virtue of the ANN architecture and automatic differentiation through the network, the functional dependence upon the first derivatives with respect to the Cartesian coordinates of the particles in the system are explicitly available, which enables propagation of biases in the CVs to forces on atoms and permits the learned CVs to be seamlessly incorporated into any off-the-shelf CV biasing enhanced sampling approach. The ``chicken-and-egg problem'' of data-driven CV discovery observes that CV discovery requires training trajectories spanning the thermally relevant phase space, and trajectories spanning the thermally relevant phase space require good CVs to drive sampling\cite{Clementi13MountainPasses}. MESA solves this problem by interleaving successive rounds of CV discovery and enhanced sampling until there is no additional round-to-round enhancement in exploration of phase space or change in number or character of the learned CVs. The first round of the CV discovery process may be initialized using short unbiased trajectories, simulated annealing, or biased trajectories in intuited or physically-motivated CVs. Subsequent rounds of the process are conducted by performing enhanced sampling in data-driven CVs discovered by the ANN. A schematic illustration of the PINES approach with parallel bias metadynamics as the CV biasing enhanced sampling approach is illustrated in Fig.\ \ref{fig:PINESflowchart}.

\subsubsection{Permutation Invariant Vector (PIV) featurization}

PINES employs a Permutation Invariant Vector (PIV) \cite{pipolo2017navigating,Pietrucci2013PIV,Pietrucci2020PIV} representation of the molecular system to permit discovery of translationally, rotationally, and permutationally-invariant CVs for enhanced sampling. The computational graph underpinning PINES is simply a PIV featurization of the system followed by a regularized autoencoder for data-driven CV discovery (Fig.\ \ref{fig:PINESflowchart}a,b). In this sense, it is very similar to MESA but employs a PIV representation of the molecular system. Importantly, gradients can be backpropagated end-to-end through the autoencoder encoder and PIV featurizer to permit calculation of the gradients in the CVs required for enhanced sampling calculations. 

The PIV featurization can be conceived as a concatenation of instantaneous pair correlation functions specifying the ordered set of pairwise distances between the chemically identical and indistinguishable atoms of each class (e.g., all water O atoms, all \ce{Na+} ions, etc.) \cite{pipolo2017navigating,Pietrucci2013PIV,Pietrucci2020PIV}. Since the featurization is constructed from pairwise distances it is naturally invariant to rigid translation and rotation, and by virtue of the inherent ordering of pairwise distances within an instantaneous pair correlation function, it is also invariant to permutations of the labels attached to the indistinguishable atoms of each class. Since the PINES network is trained to reconstruct the PIV representation of the system, the learned CVs are also inherently invariant to translation, rotation, and permutation. Importantly, the PIV representation of a system is complete in the sense that if all pairwise distances are incorporated into the PIV featurization, rigid graph theory dictates that the Cartesian coordinates of the particles constituting the system can be unambiguously reconstructed up to translation, rotation, and permutation \cite{Singer2008RigidGraph,dokmanic2015euclidean}. 

\begin{figure}[ht!]
\centering
  \includegraphics[width=1\textwidth]{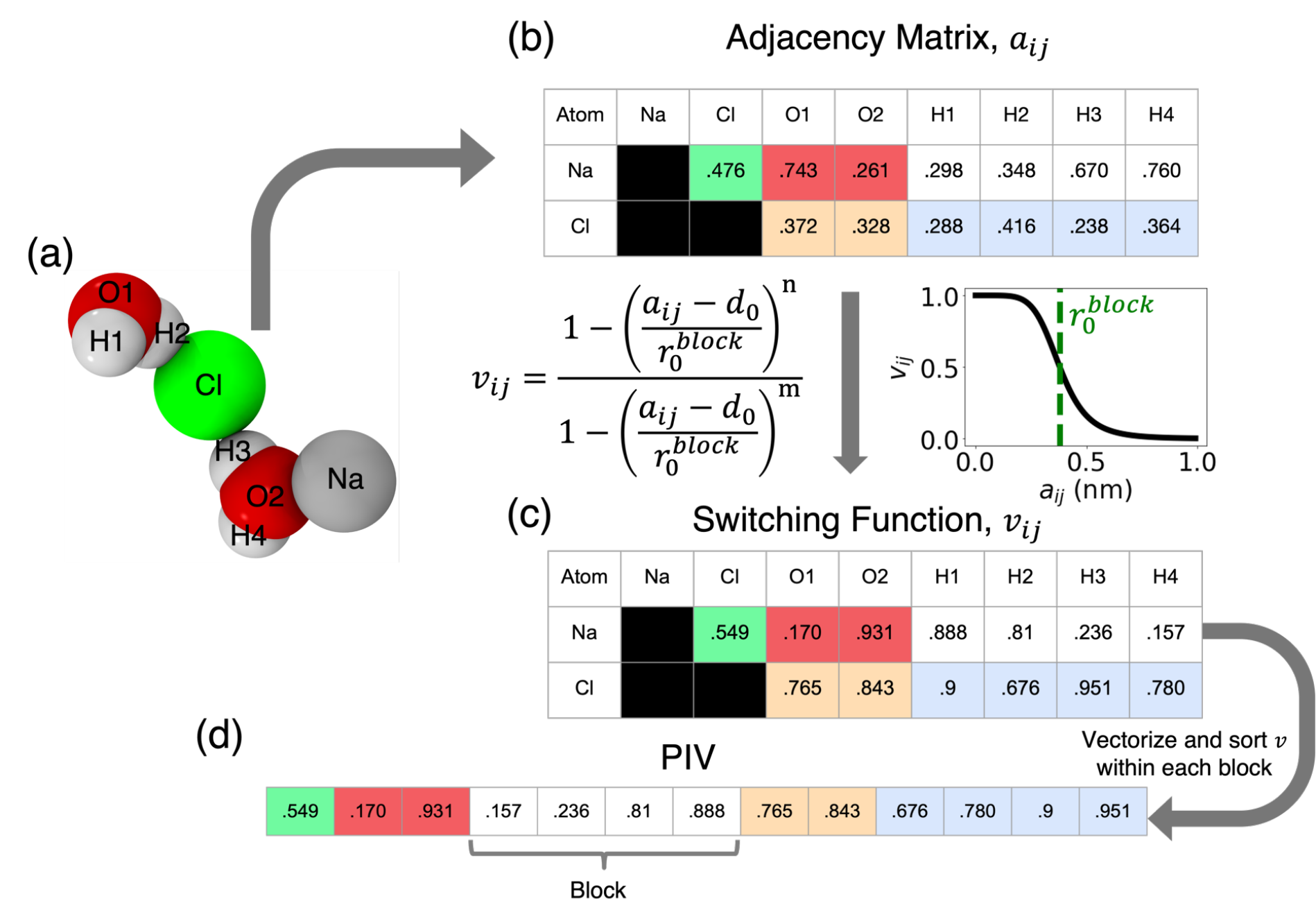}
  \caption{Schematic illustration of calculation of the PIV featurization of a \ce{NaCl} ion pair in water. (a) Snapshot of an instantaneous state of the \ce{NaCl} ion pair and the two most proximate water molecules. All molecular renderings in this work are constructed using Visual Molecular Dynamics (VMD) \cite{VMD}. (b) We convert the Cartesian coordinates of the atomic system into an adjacency matrix recording all pairwise distances $a_{ij}$ between all pairs of atoms $i$ and $j$. The symmetry of the adjacency matrix means we need only record the upper triangle and is arranged into a block structure by atom class such that indistinguishable atoms are grouped together. Since the number of pairwise distances increases quadratically with the number of atoms, we may also choose to restrict the atoms over which we construct the adjacency matrix by, for example, omitting all but the alpha-carbons in a protein or disregarding the far-field solvent molecules around a solute. In the interests of space, we present here the abridged adjacency matrix where we have colored each block or group of (indistinguishable) pairwise distances: Na-Cl (green), Na-O (red), Na-H (white), Cl-O (orange), and Cl-H (blue). (c) The elements of the adjacency matrix are transformed by a rational switching function, the parameters of which are tuned to present the largest dynamic range and sensitivity to configurational changes in the short- to medium-ranged field. It also has the desirable property of scaling the pairwise distances to a $(0,1)$ range. We choose $r_{0}$ in each block ($r_{0}^{block}$ ) to be the mean of the first and second peaks of the RDF, set $d_{0}=0$ and $m = 2n$, and tune $n$ such that $v_{ij} \approx  0.9$ at the $a_{ij}$ value of the first peak and $v_{ij} \approx  0.1$ at the $a_{ij}$ value of the second peak in the RDF. (d) The elements within each block of the switched adjacency matrix are placed in non-descending order and concatenated to form PIV.}
  \label{fig:pipeline}
\end{figure}

PIVs were introduced by Pietrucci and co-workers \cite{pipolo2017navigating,Pietrucci2013PIV,Pietrucci2020PIV} and have been used, among other applications, to efficiently drive sampling of different phases of water \cite{pipolo2017navigating}. We illustrate in Fig.\ \ref{fig:pipeline} the construction of a PIV featurization for a \ce{NaCl} ion pair in water. The PIV featurization proceeds in three steps. First, we compute the molecular adjacency matrix with elements $a_{ij}$ recording the Euclidean pairwise distances between atoms $i$ and $j$. We arrange the elements of the matrix into a block structure where each block contains the set of pairwise distances between two atom classes (i.e., groups of chemically indistinguishable atoms). Within each block the ensemble of pairwise distances are unchanged under permutational relabeling of the chemically indistinguishable atoms within each class. Second, we pass the pairwise distances through a rational switching function to produce the switched adjacency matrix,
\begin{equation}
v_{ij} = \frac{1-\left( \frac{a_{ij} - d_{0}}{r_{0}} \right)^n}{1-\left( \frac{a_{ij} - d_{0}}{r_{0}} \right)^m}
\end{equation}
where the hyperparameter $d_{0}$ is minimum interparticle distance, $r_{0}$ is the interparticle distance which lies on the midpoint of the switching function curve (Fig.~\ref{fig:pipeline}b), and $n$ and $m$ are non-negative numbers which control the steepness of the switching function curve and thus determine the resolution of the various pairwise interactions. This transformation has the two desirable effects of scaling the distances to the interval $(0,1)$ and producing values close to unity for short distances, values close to zero for large distances, and a switching range between these two limits over which the transformation has the most sensitivity to changes in $a_{ij}$. This ensures that the $v_{ij}$ values passed as input to the encoder are appropriately scaled to $(0,1)$ range and, by tuning the parameters of the switching function, permits us to focus the dynamic range of $v_{ij}$ upon the physical range of distances over which we wish the network to pay the most attention. Typically, we set the parameters based on a radial distribution function (RDF) collected over a short unbiased trial simulation of the system in its stable state (Fig.~S1). We choose $r_{0}$ to be the mean of the first and second peaks of the RDF, set $d_{0}=0$ and $m = 2n$ 
and tune $n$ such that $v_{ij} \approx  0.9$ at the $a_{ij}$ value of the first peak and $v_{ij} \approx  0.1$ at the $a_{ij}$ value of the second peak in RDF. This ensures that the $v_{ij}$ values are most responsive to particles moving between the first two shells of their radial distribution where it is typically most important to highlight structural differences between various system configurations and become insensitive in the far field \cite{Pietrucci2020PIV}. When appropriately tuned in this fashion, the PIV featurization show large inter-class differences between qualitatively different system configurations and small intra-class differences for configurations in the same structural class \cite{Pietrucci2020PIV}. Third, we sort the $v_{ij}$ values within each block of the switched adjacency matrix in non-descending order and concatenate them into a single vector -- the eponymous PIV. The sorting operation is the critical step in introducing permutational invariance. 


Importantly, construction of the PIV from the Cartesian coordinates of the system is a deterministic and differentiable operation that permits backpropagation of gradients from the learned CVs to the PIV featurization and then the Cartesian coordinates by automatic differentiation. A well known deficiency of PIVs, and one that is shared by all featurizations that achieve permutational invariance by ordering, is that the featurization can exhibit kinks and its derivative therefore possess discontinuities at locations in the configurational phase space where the ordering of the particles changes \cite{musil2021physics}. Empirically, we encountered no difficulties or instabilities in either our CV learning process or our enhanced sampling calculations in the learned CVs. It is possible that the discontinuities are sufficiently gentle and/or rare that they do not present any difficulties, but this issue may warrant additional study to identify situations where these singularities may present difficulties.

\subsubsection{Autoencoder network design and training}

The second component of PINES is an autoencoding neural network that performs unsupervised nonlinear dimensionality reduction of molecular simulation training trajectories to discover low-dimensional embeddings into a small number of high-variance CVs \cite{kingma2013auto,kingma2019introduction,Ferguson2018MESA,Ferguson2018CVDiscovery}. In this work, we employ a maximum mean discrepancy-Wasserstein autoencoder (MMD-WAE)\cite{Ermon17InfoVAE}. The network is trained by minimizing the loss function comprising a mean squared error (MSE) term to encourage high fidelity reconstruction and a maximum mean discrepancy (MMD) loss to regularize the latent space,
\begin{equation}
\mathcal{L} = \frac{1}{N}\sum_{i=1}^{N}(PIV_{i} - \widehat{PIV}_{i})^2 + E_{P}[k(x,x^\prime)] + E_{Q}[k(y,y^\prime)] - 2E_{P,Q}[k(x,y)],
\end{equation}
where $k(z,z^\prime) = \exp{(z-z^\prime)^2}$ is a Gaussian kernel with unit bandwidth, $\frac{1}{N}\sum_{i=1}^{N}(PIV_{i} - \widehat{PIV}_{i})^2$ is the element-wise mean squared error of the reconstruction, $E_{P}[k(x,x^\prime)] + E_{Q}[k(y,y^\prime)] - 2E_{P,Q}[k(x,y)]$ is the MMD loss, $P(x)$ is the probability distribution of the latent space, and $Q(y)$ is the target latent space distribution. The MMD loss is equal to zero only when the two distributions are identical. In this work, we choose the target distribution to be a normal distribution with a zero mean and unit variance.

We note that in principle it may be advantageous to sort the elements within each block of the reconstructed PIV into non-descending order prior to calculating the MSE loss. In this way the decoder need only learn the distribution of values within each block instead of also being required to learn the sorting rule. Empirically, we find that the decoder is able to readily learn to produce values in non-descending order and only rarely violates proper sorting within the reconstructed PIV. It appears, therefore, that pre-loss sorting of the reconstructed PIV is unnecessary. Additionally we draw attention to the possibility of using a non-uniform weights over the various contributions to the MSE term in order to modulate the importance of reconstruction fidelity for particular interactions in the system \cite{Pietrucci2020PIV}, although in our work we observed satisfactory performance using uniform weighting.

The encoder and decoder are both implemented as fully-connected feedforward neural networks with three hidden layers (Fig.\ \ref{fig:PINESflowchart}b). For all systems considered in this work, the overall network architecture is $M$-64-32-16-$d$-16-32-64-$M$, where $M$ is the dimensionality of the PIV featurization and $d$ is the number of nodes in the bottleneck layer that defines the number of CVs to be discovered. An appropriate value for $d$ is determined each round by training networks with $d$ ranging from 1-8 and selecting the appropriate value of $d$ from a knee in a plot of the fraction of variance explained curve using the L-method of Salvador and Chan \cite{Ferguson2018MESA,Ferguson2018CVDiscovery,Chan:L-method} (Fig.~S2). We employ $tanh$ activation functions between all hidden layers and a sigmoid activation in the final layer of the decoder. Batch normalization is employed between all hidden layers except the layer preceding the reconstructed PIV, which we find to be critical in avoiding saturation and stabilizing CV discovery. We construct and train the network using PyTorch\cite{PyTorch} with PyTorch Lightning\cite{falcon2020pytorchlightning} and train it using backpropagation employing the Adam optimizer\cite{AdamOptimation} with a batch size of 100 and learning rate of $10^{-4}$. Training was terminated when the validation loss stopped decreasing employing a patience of 10 epochs. Training was performed on a NVIDIA Quadro RTX 6000 GPU card and, for the systems studied in this work, typically required hundreds of epochs and tens of GPU-minutes. After training is complete, we discard the decoder and pass the output of the bottleneck layer through a principal components analysis (PCA) whitening transformation (Fig.\ \ref{fig:PINESflowchart}b) \cite{shlens2014tutorial}. We find this step useful in disentangling the learned CVs by imposing ordering by variance that helps to both stabilize the round-to-round discovery process and also improves enhanced sampling by orienting the biasing potentials along the orthogonalized PCA subspace. This transformation can be thought of as a simple change of basis in which the dimensionality of the latent space is conserved and the whitened CVs represent a rotation of the manifold to align the axes with the highest variance regions of the embedding. 


\subsubsection{Enchanced sampling along discovered CVs using parallel bias metadynamics}\label{sec:PBMetaD}

The learned CVs within each round are used to conduct enhanced sampling of configurational space using parallel bias metadynamics (PBMetaD)\cite{Pfaendtner2015PBMD}. The PINES approach is compatible with any off-the-shelf CV biasing enhanced sampling technique, but we elect to use PBMetaD due to its efficient scaling to large numbers of CVs by the application of multiple simultaneous one-dimensional biases (Fig.\ \ref{fig:PINESflowchart}c). 
In this work, we never have need to drive sampling more than three CVs, but the PBMetaD approach has been demonstrated up to several tens of CVs \cite{prakash2018biasing}. We couple all learned CVs to a well-tempered variant of PBMetaD enhanced sampling calculation implemented within the PLUMED2 library \cite{PLUMED2,Parrinello2008WTMD,Pfaendtner2015PBMD}. The hyperparameters for the PBMetaD calculations are detailed for each of the three systems in Section \ref{sec:MD}.

Convergence of the PBMetaD calculations are assessed using two criteria \cite{bussi2020using}. First, we check that the biasing potential has become quasi-static by assuring that the height of the Gaussian biases deposited over the course of the simulation drop below a particular fraction of the initial height. In early iterations we adopt a less stringent criterion of 10\% but in later rounds, we refine this to a more strict criterion of 1\%. Second, we monitor the distribution of the CVs and check for diffusive sampling over the full range explored by each CV to verify that the simulation can freely diffuse over the barriers in all coupled coordinates. 

\subsubsection{Convergence assessment}\label{sec:convergence}

Convergence of PINES is assessed by measuring the dimensionality $d$ and cosine similarity $\rho$ of the CVs learned between successive iterations and the expansion of the explored phase space. Stabilization of $d$ of the MMD-WAE bottleneck layer indicates that no additional dimensions are being explored in successive biasing rounds, a cosine similarity of $\rho$ $>$ 0.9 between the CVs discovered in two consecutive rounds -- computed by projecting the present round's simulation data into the CVs learned in the present and previous rounds -- is used to assess whether the CVs themselves are approximately constant between successive rounds, and we verify that the boundary of explored phase space is no longer expanding (Fig.\ \ref{fig:PINESflowchart}d). 

Once all convergence criteria are satisfied, we declare the PINES procedure to be complete. At this point we construct a training data set by collating the biased trajectories from all prior rounds and use this concatenated data set to run a final round of CV discovery and PBMetaD enhanced sampling. For large datasets, we perform uniform sampling across the multidimensional CV space used for the latest round of biasing to generate a smaller, subsampled dataset that is representative of the whole CV space but more amenable to training a terminal network to identify the optimal embedding. These final CVs are used to perform a PBMetaD production run to demonstrate their impact on accelerating exploration of phase space. The unbiased free energy surface is recovered from the terminal PBMetaD biased trajectories by applying the standard Torrie-Valleau sampling weights to each configuration to eliminate the effect of the applied bias and enable projection into arbitrary CVs beyond those in which the biased sampling was conducted \cite{Pfaendtner2015PBMD,Valleau:1977:Umbrella}. Empirically, we do not observe any significant benefit from applying this reweighting to the biased trajectories in each individual CV discovery round, but this, of course, is possible \cite{Stolz2021FEBILAE}.

\subsection{All-atom molecular dynamics simulations}\label{sec:MD}

All-atom molecular dynamics simulations of each of the three systems considered in this work were performed using the GROMACS 2021.1 simulation suite \cite{abraham2015gromacs} patched with an in-house modified version of the PLUMED2 v2.8.0 enhanced sampling libraries \cite{PLUMED2}. This patch implements the PINES approach by integrating the newly developed module named PINES and the ANN module previously contributed by Chen and Ferguson\cite{Ferguson2018MESA}. We developed the PINES module in PLUMED2 using the PIV module previously contributed by Pipolo and Pietrucci\cite{Pietrucci2013PIV,Pietrucci:17:PRL} as a template. In the ANN module, we added support for batch normalization. Biasing of the translationally, rotationally, and permutationally-invariant CVs discovered by PINES can be performed by any off-the-shelf CV biasing enhanced sampling technique. In this work, we use the PBMetaD PLUMED2 module contributed by Bonomi and Pfaendtner\cite{Pfaendtner2015PBMD} in PLUMED2 for enhanced sampling calculations. The new PINES and modified ANN modules along with a helper bash script for patching with PLUMED2 are available at \url{https://github.com/Ferg-Lab/pines.git}, and will be submitted for incorporation into a future release of PLUMED2\cite{PLUMED2}. All the data and PLUMED input files required to perform the molecular dynamics simulations of the three systems have been hosted on PLUMED-NEST, the public repository of the PLUMED consortium \cite{Bonomi2019}, under project ID \texttt{plumID:23.024} (\url{https://www.plumed-nest.org/eggs/23/024/}).
 
\subsubsection{13-particle Ar cluster}
We consider a system of 13 Ar atoms to demonstrate the utility of PINES in driving assembly and disassembly of a simple Lennard-Jones cluster of identical particles\cite{Doye:LJ13:Web}. The Ar atoms were modeled in vacuum using Lennard-Jones non-bonded interactions with $\sigma=0.34$ nm and $\epsilon=1$ kJ mol$^{-1}$.\cite{dawid1997interaction, briant1975molecular}. Particles were randomly placed in a 2.5$\times$2.5$\times$2.5 nm$^{3}$ cubic box implementing periodic boundary conditions. The system was subjected to steepest descent energy minimization to eliminate forces larger than 1 kJ mol$^{-1}$ nm$^{-1}$. After energy minimization, a short equilibration simulation was performed in a NVT ensemble for 100 ps. The classical mechanical equations of motion were integrated using the leap-frog algorithm\cite{Hockney:74:JPhysChem} with a 2 fs time step. Particle velocities were initialized according to the Maxwell-Boltzmann distribution at 50 K. Simulations were conducted in the NVT ensemble in which temperature is maintained using a velocity-rescaling thermostat\cite{Bussi:07:JChemPhys} employing a time constant of 0.1 ps. Lennard-Jones interactions were smoothly shifted to zero at a cutoff of 1.0 nm. The unbiased production run for the initial PINES round was conducted for 25 ns and frames were saved at a frequency of 1 ps for training the autoencoder in the subsequent PINES round. We conducted 25 ns of PBMetaD simulations in subsequent PINES iterations at a temperature of 50 K employing an initial Gaussian height of $W$ = 0.5 kJ mol$^{-1}$, a Gaussian width of $\sigma$ = 0.1 for each discovered CV, a bias factor of $\gamma$ = 10, and deposition pace of 500 steps (1 ps). The PIV featurization contained one Ar-Ar interaction block with 78 elements corresponding to all possible pairwise distances. Simulations were performed on 5 $\times$ 2.40 GHz Intel Xeon Gold 6148 CPU cores and 1 $\times$ NVIDIA TITAN V GPU, achieving execution speeds of $\sim$120 ns per day.

\subsubsection{NaCl ion pair in water}
An NaCl ion pair in water was used as a classic test system to assess the capacity of PINES to discover solvent-inclusive CVs governing the association/dissociation of the ion pair and use these to efficiently accelerate sampling\cite{jung2019artificial,mullen2014transmission,kellermeier2016entropy,geissler1999kinetic,wang2022influence} We utilized the Amber99SB-ILDN forcefield\cite{lindorff2010Amber99SBILDN} with Joung-Cheatham\cite{joung2009molecular,joung2008determination} ion parameters. The \ce{Na+} and \ce{Cl-} ions were solvated to a density of 1 g cm\textsuperscript{-3} with 508 TIP3P\cite{jorgensen1983comparison} water molecules in a cubic simulation box of size 2.5$\times$2.5$\times$2.5 nm$^3$. A cutoff of 1 nm was applied for the Lennard-Jones interactions. Long-range electrostatics were treated by particle mesh Ewald summation\cite{Darden:93:JChemPhys,essmann1995smooth} employing a real-space cutoff of 1.0 and 0.16 nm Fourier grid spacing that were optimized during runtime. The neighbor list was generated using the Verlet method and was updated every 10 steps (20 fs). An energy minimization procedure with steepest-descent algorithm was applied to relax the system after solvation by adjusting the configuration to have forces between particles with less than 1 kJ mol$^{-1}$ nm$^{-1}$. After energy minimization, a short equilibration simulation was performed in a NVT ensemble for 100 ps. The leapfrog algorithm\cite{Hockney:74:JPhysChem} with a timestep of 2 fs was applied for propagating the equations of motion in the simulations. The initial velocities for the equilibration run were generated using the Maxwell-Boltzmann distribution at 300 K. A velocity-rescale thermostat\cite{Bussi:07:JChemPhys} was applied for controlling the temperature during the equilibration run. For the production run, the final configuration and velocities from the equilibration run were used to initiate the simulation. Frames at a frequency of 2 ps are saved from the production run for training the autoencoder in the subsequent PINES rounds. The PIV featurization contained five interaction blocks with a total of 61 elements distributed across the blocks as 1 Na-Cl, 10 Na-O, 20 Na-H, 10 Cl-O, and 20 Cl-H elements.

The initial production unbiased simulation was performed for 50 ns. For PBMetaD in subsequent rounds, we set the initial gaussian height as $W$ = 1.2 kJ/mol, a Gaussian width of $\sigma = 0.1$ for each of the discovered CVs, a bias factor of $\gamma$ = 20, and a deposition pace of 500 timesteps (1 ps). Each simulation in the PINES rounds with enhanced sampling along discovered CVs using PBMetaD was initially performed for 15 ns in a NVT ensemble at 300 K but was extended until all the convergence metrics for PBMetaD were satisfied. Simulations were performed on 5 $\times$ 2.40 GHz Intel Xeon Gold 6148 CPU cores and 1 $\times$ NVIDIA TITAN V GPU, achieving execution speeds of $\sim$50 ns per day.

\subsubsection{\texorpdfstring{Aqueously solvated \textit{n}-pentatetracontane (\ce{C_{45}H_{92}}) polymer chain}{Aqueously solvated n-pentatetracontane (C45H92) polymer chain}}
An aqueously solvated \textit{n}-pentatetracontane (\ce{C_{45}H_{92}}) polymer chain was used to demonstrate the ability of PINES to learn solvent-inclusive CVs for hydrophobic collapse and use these CVs to enhance sampling of the elongation/collapse transition\cite{Patel2021PolymerSolvationINDUS}. The initial structure of the \textit{n}-pentatetracontane alkane chain was modeled using the TraPPE-UA\cite{martin1998transferable,eggimann2014online} potential that models the chain as a string of \ce{CH3}/\ce{CH2} united atoms. The system was solvated with 11,188 SPCE\cite{berendsen1987missing} water molecules in a cubic simulation box of size 7.0$\times$7.0$\times$7.0 nm$^3$. A cutoff of 1 nm was applied for the Lennard-Jones interactions. Long-range electrostatic interactions are handled by particle mesh Ewald method \cite{Darden:93:JChemPhys,essmann1995smooth} employing a real-space cutoff of 1.0 and 0.16 nm Fourier grid spacing that were optimized during runtime. The neighbor list was generated using the Verlet method\cite{verlet1967computer, pall2013flexible} and is updated every 20 steps (40 fs). An energy minimization procedure with steepest-descent algorithm was applied to relax the system after solvation by adjusting the configuration to have forces between particles with less than 1 kJ mol$^{-1}$ nm$^{-1}$. After energy minimization, a short equilibration simulation was performed in a NVT ensemble for 100 ps. The leapfrog algorithm\cite{Hockney:74:JPhysChem} with a timestep of 2 fs was applied for propagating the equations of motion. The initial velocities for the equilibration run were generated using the Maxwell-Boltzmann distribution at 300 K. A velocity-rescale thermostat\cite{Bussi:07:JChemPhys} was applied for controlling the temperature during the equilibration run. For the production run, the final configuration and velocities from the equilibration run were used to initiate the simulation. The unbiased production run in the initial PINES round was conducted for 100 ns with a timestep of 2 fs. Frames at a frequency of 2 ps are saved from the production run for training the autoencoder in the subsequent PINES rounds. The PIV featurization contained 45 interaction blocks with a total of 108 elements distributed across the blocks. The blocks are characterized by the intramolecular interactions between nine evenly spaced \ce{CH2} united atoms from position 3 to 43 in the chain, yielding 36 blocks each containing only a single element, as well as the intermolecular interactions with their respective eight closest water oxygens, yielding nine blocks each containing eight elements. We note that the \ce{CH2}-\ce{CH2} united atom blocks could be represented as 18 blocks of two elements to mod out the symmetry associated with the head-to-tail inversion symmetry of the alkane chain, however we choose not to do this here.

For the PBMetaD calculations, we set the initial gaussian height as $W$ = 1.2 kJ mol$^{-1}$, a Gaussian width of $\sigma = 0.1$ for each discovered CV, a bias factor of $\gamma$ = 20, and a deposition pace of 500 timesteps (1 ps). Each simulation in the PINES rounds with enhanced sampling along discovered CVs using PBMetaD was initially performed for 30 ns in a NVT ensemble at 300 K but was extended until all the convergence metrics for PBMetaD were satisfied. Simulations were performed on 5 $\times$ 2.40 GHz Intel Xeon Gold 6148 CPU cores and 1 $\times$ NVIDIA TITAN V GPU, achieving execution speeds of $\sim$10 ns per day.

\section{\label{sec:results}Results and Discussion}

We demonstrate the general utility and applicability of PINES to discover translationally, rotationally, and permutationally invariant CVs for enhanced sampling in applications to three molecular systems: (i) self-assembly of a 13-atom \ce{Ar} cluster, (ii) assembly/disassembly of a \ce{NaCl} ion pair in water, and (iii) hydrophobic collapse of a \ce{C_{45}H_{92}} $n$-pentatetracontane polymer chain.

\subsection{Self-assembly of a 13-particle Ar cluster}

We first demonstrate the application of PINES to the self-assembly of a 13-particle Ar cluster in vacuum. We hypothesize that PINES can learn permutationally-invariant multi-body collective variables associated with the assembly and disassembly of the identical Ar atoms between solid-like, liquid-like, and vapor-like phases. After some trial-and-improvement exploration, we chose to perform the simulations at a temperature of 50 K, which we observed to provide good unbiased sampling of the solid, liquid, and vapor-like phases and therefore access to an unbiased ground truth against which to compare the biased sampling calculations under PINES. As in our previous work with MESA, we determine the dimensionality of the latent space by training independent networks with 1-8 nodes in the bottleneck layer of autoencoder and utilizing the L-method by Salvador and Chan\cite{Chan:L-method} to identify a knee in the fraction of variance explained (FVE) curve (Fig.~S2). With this method, we identified a latent space of dimensionality $d$=3. 

\begin{figure}[ht!]
\centering
  \includegraphics[width=0.95\textwidth]{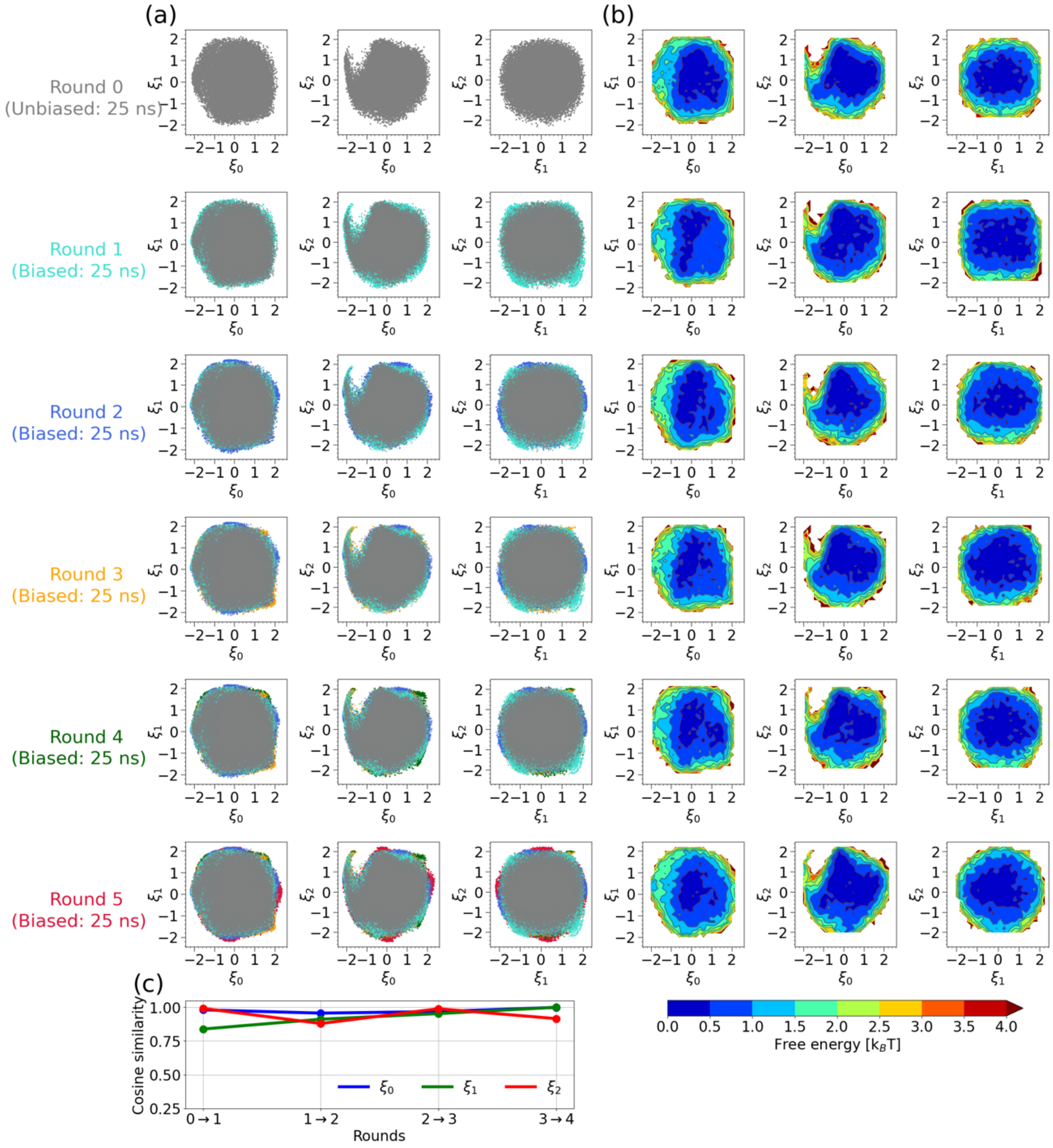}
   \caption{Application of PINES to a 13-particle Ar system in vacuum at 50 K. (a) Each column shows a 2D embedding of the round-by-round accrued sampling projected into the converged CVs $\xi_{0}$, $\xi_{1}$, and $\xi_{2}$ at the terminal Round 5. The unbiased simulation quite thoroughly samples the thermally accessible phase space, yet each biased round advances the frontier into unsampled regions on the periphery of the explored region. (b) Corresponding free energy surfaces (FES) computed by reweighting of the biased simulation data collected in the current round of sampling. (c) Cosine similarity between corresponding CVs in successive rounds. The high cosine similarity value indicates rapid convergence of the iterative CV discovery and enhanced sampling cycles within the five rounds of PINES.}
   \label{Fig3}
\end{figure}

In Fig.~\ref{Fig3}a we show the round-by-round sampling as 2D projections along the three CVs $\xi_{0}$, $\xi_{1}$, and $\xi_{2}$ learned from the terminal PINES network. The initial unbiased simulation used to seed PINES (gray) provided a good initial coverage of the accessible configurational space and each subsequent biased round (different colors) extends the frontier into new unsampled regions of phase space around the periphery of the explored region and probes higher free energy configurations. We present in Fig.~\ref{Fig3}b the reweighted (i.e., unbiased) free energy surfaces (FES) associated with each sampling round that better expose the variations in sampling density over the three CVs. In particular, we observe that the biased sampling identifies a low-$\xi_0$ metastable minimum that was not resolved by the unbiased calculations in Round 0. As we discuss below, this minimum at $(\xi_0 \approx -2, \xi_1 \approx 0, \xi_2 \approx 1)$ corresponds to a rarefied vapor-like state that is incompletely explored in the unbiased calculations.

We assess convergence of the iterative cycles of CV learning and biased sampling by computing the cosine similarity between corresponding CVs in successive rounds. As illustrated in Fig.~\ref{Fig3}c, we find each of the three learned CVs to quickly stabilize with round-to-round cosine similarity scores in excess of $\rho$ = 0.9 for Round 1 and above. Together with the relatively modest expansion of the configurational phase space (Fig.~\ref{Fig3}a) and stabilization of the FES (Fig.~\ref{Fig3}b), we declare the PINES procedure to be converged within the five rounds.

\begin{figure}[ht!]
\centering
  \includegraphics[width=1\textwidth]{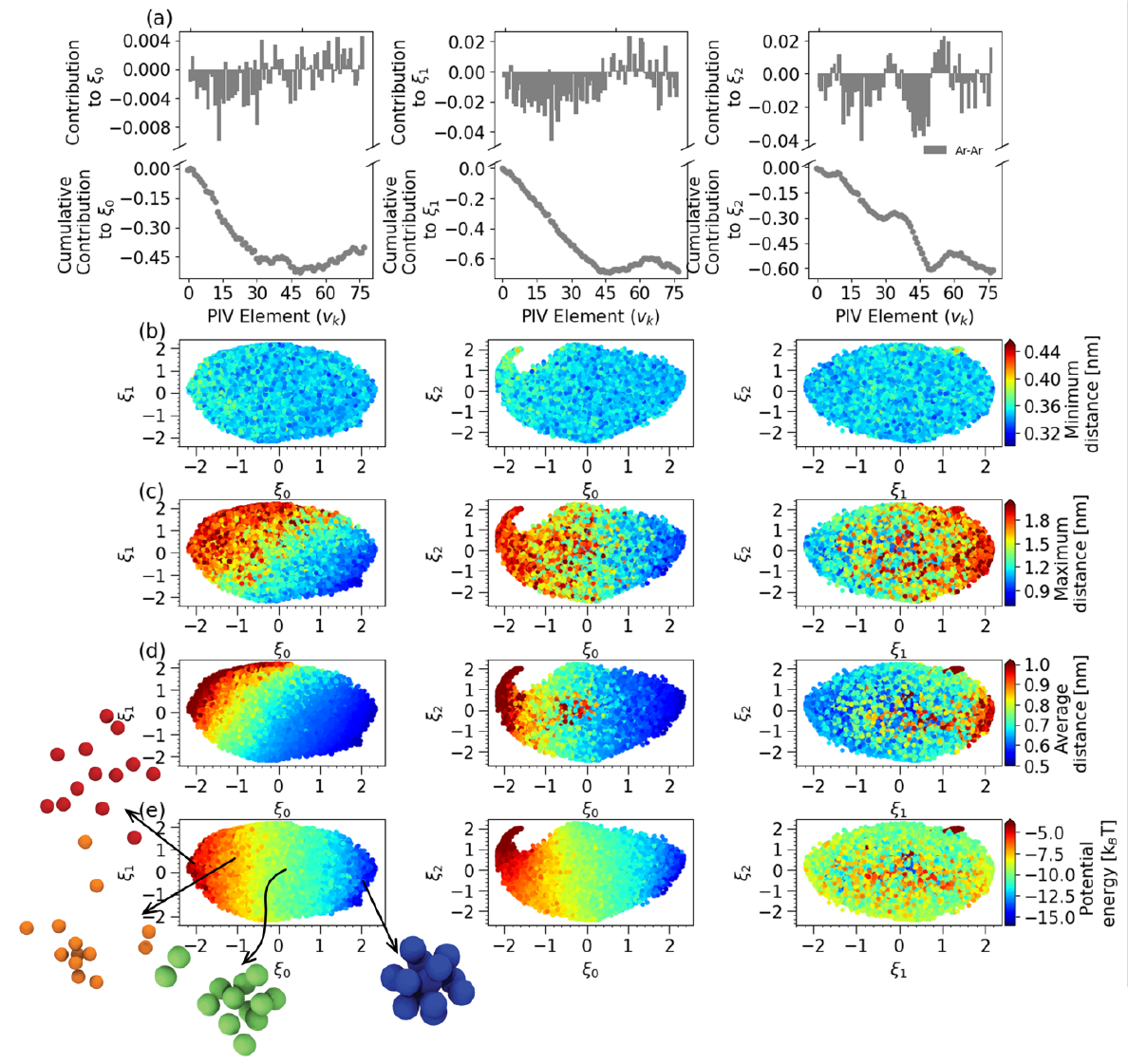}
  \caption{Bottom-up and top-down analysis of terminal learned CVs for the 13-particle Ar system. (a) Split plots show the signed contribution from each PIV element to each CV $\{ \xi_0, \xi_1, \xi_2 \}$ as a bar chart on top and as a cumulative sum of the signed contributions on the bottom. The magnitude of contributions on the bar chart emphasize the importance of individual PIV elements while the slopes in different regions of the scatter plot illuminate the concerted effect of atom groups on each CV. (b-e) Projections of the terminal Round 5 sampling data into the terminal CVs and colored by physically interpretable candidate variables. The potential energy, which is known a priori to be the putative CV for a Lennard-Jones fluid, correlates well with $\xi_{0}$. Selected snapshots are presented to show the smooth transition across $\xi_{0}$ from the gas-like phase (red, orange) to the condensed phase (green), and finally to the highly ordered solid-like phase (blue). The gradients in $\xi_{0}$ and $\xi_{1}$ with respect to the average and maximum pairwise distances indicates that PINES has learned these as important variables in driving assembly/disassembly.}
  \label{Fig4}
\end{figure}

It is informative to analyze the terminal learned CVs to provide insight about the high variance emergent collective modes within the system. By extracting the weights of the trained encoder, we perform layerwise matrix multiplication to compute a single scalar weight encapsulating the extent to which each PIV element contributes to each of the learned CVs integrated over all possible pathways through the network. Although this analysis does not account for correlations between the various PIV elements, it is useful in providing a gross assessment of how strongly each permutationally invariant pairwise distances influences each of the learned CVs. We present in Fig.~\ref{Fig4}a a graphical illustration of the contribution of each PIV element $v_{k}$ to each of the the terminal learned CVs $\{ \xi_0, \xi_1, \xi_2 \}$, and also a cumulative sum over these weights indicating the net contribution and cooperative impact over multiple PIV elements at increasing pairwise distances. The absence of strong outliers in the bar charts indicates that none of the ${13 \choose 2}$ = 78 PIV elements substantially dominate the values of the three learned CVs. A similar trend is manifested in the cumulative sums that indicate an approximately linear trend indicative of similar magnitude and similar signed contributions out to PIV element 35 or so. Recalling that the pairwise distances are passed through a switching function (Fig.~\ref{fig:pipeline}) the larger pairwise distances appear earlier in the PIV, and for the shorter range interactions beyond element 35, we see a more interesting and challenging to interpret structure in the sign and magnitude of the PIV weights. The identical nature of all particles in this system and necessarily cooperative nature of assembly and disassembly is consistent with the absence of any single dominant contribution from the PIV.

We complement this bottom-up analysis of the learned CVs with a top-down interrogation in which we color projections of the Round 5 simulation data into the terminal CVs by candidate physical order parameters (Fig.~\ref{Fig4}b-e). In doing so, we find $\xi_{0}$ to be strongly correlated with the potential energy of the system (Fig.~\ref{Fig4}e). Since the potential energy is a sum over pairwise interactions, this makes $\xi_0$ a good correlate for the degree of association of the 13-particle cluster, with the compact, low-potential energy solid-like state residing at high $\xi_0$, the dissociated, high-potential energy vapor-like state at low $\xi_0$, and the liquid-vapor-like, partially associated / partially dissociated states at intermediate $\xi_0$. The average pairwise distance (Fig.~\ref{Fig4}d) and maximum pairwise distance (Fig.~\ref{Fig4}c) are both well-correlated with a linear combination of $\xi_{0}$ and $\xi_{1}$, with the ``red hook'' apparent in the $\xi_0$-$\xi_2$ projections containing the most highly dissociated and rarefied vapor-like configurations. The minimum pairwise distance fails to show strong correlations with any of the three CVs (Fig.~\ref{Fig4}b). In sum, the three permutationally-invariant CVs learned by PINES as important collective modes to drive sampling of the configurational exploration of the system reflect the potential energy, compactness, and average/maximum pairwise interatomic distance.

\begin{figure}[ht!]
\centering
  \includegraphics[width=1\textwidth]{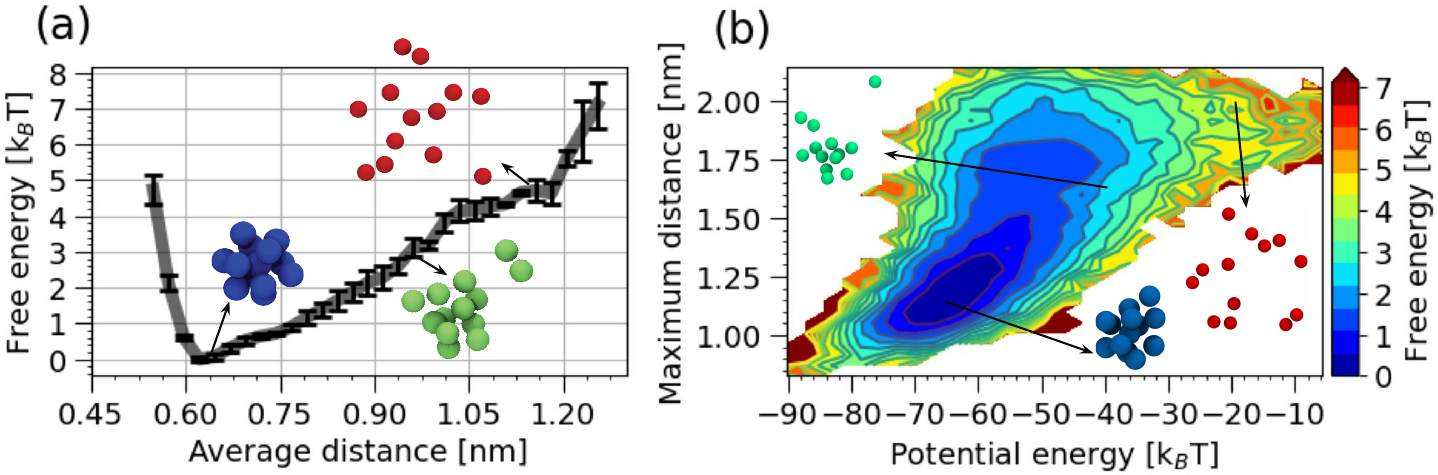}
  \caption{Reweighted free energy surfaces from the terminal round of enhanced sampling for the 13-particle Ar system projected into physically interpretable order parameters. (a) A 1D FES projected into the average pairwise interatomic distance. Representative snapshots show that the global free energy minimum to be occupied by a solid-like, filled icosahedron solid with an average pairwise distance of $0.62$ nm. (b) A 2D FES projected into the potential energy and maximum pairwise interatomic distance. Representative snapshots illustrate the solid-like global free energy minimum and liquid-like and vapor-like phases.}
  \label{Fig5}
\end{figure}

Finally, in Fig.~\ref{Fig5} we present our reweighted FES collected from our final round of enhanced sampling in the converged terminal learned CVs. We emphasize that these data were collected from enhanced sampling calculations conducted in the learned CVs (cf.\ Fig.~\ref{Fig3}), but that after sampling is complete, we are at liberty to reweight and project these data into arbitrary order parameters of our choosing for the purposes of expository illustration. The 1D FES constructed in the average pairwise distance clearly exposes the solid-like, filled icosahedron as the global free energy minimum of the system at an average distance of 0.62 nm (Fig.~\ref{Fig5}a). At higher average distances we enter into a liquid/vapor-like in which some fraction of the particles in the system are associated, and then terminate in a vapor-like phase with all particles dispersed lying within a weak free energy minimum at 1.17 nm approximately 4 $k_B T$ higher in free energy than the solid-like phase. The 2D FES constructed in the potential energy and maximum pairwise distance better illustrates the expansion of accessible phase space for the more disordered liquid-like and vapor-like states compared to the solid-like filled icosahdron (Fig.~\ref{Fig5}b).

The location and relative stability of the solid, liquid, and vapor-like phases is a strong function of the box size and temperature. In particular, the 2.5$\times$2.5$\times$2.5 nm\textsuperscript{3} box places strict finite size effects on the maximum possible pairwise distances available to the particles in the vapor-like phase and limits the phase space volume accessible to it. The temperature also changes the relative balance of energetic and entropic contributions and modulating this away from the 50 K temperature at which our simulations were conducted can strongly modify the underlying free energy surface. To test the transferability of the converged CVs learned by PINES in a 2.5$\times$2.5$\times$2.5 nm\textsuperscript{3} at 50 K, we conducted enhanced sampling calculations in these CVs to recover the FES in an 8-fold larger 5$\times$5$\times$5 nm\textsuperscript{3} simulation box (Fig.~S3) and at temperatures of 25 K and 100 K (Fig.~S4). Pleasingly, the CVs learned by PINES at 2.5$\times$2.5$\times$2.5 nm\textsuperscript{3} and 50 K were capable of efficient exploration of the thermally-accessible configurational phase under these new conditions without any additional retraining. In particular, the enhanced sampling calculations had no difficulty in sampling the thermodynamically unfavorable solid-like phase in the large 5$\times$5$\times$5 nm\textsuperscript{3} box (Fig.~S4) or in traversing the $\sim$40 $k_B T$ difference between the solid-like and vapor-like phases in the 25 K low temperature calculations (Fig.~S4).

Taken together, this application of PINES to a simple 13-particle Ar system demonstrates that the method is capable of learning translationally, rotationally, and permutationally invariant CVs, using these CVs to efficiently drive PBMetaD enhanced sampling calculations, and furnishing transferable order parameters capable of driving sampling at other thermodynamic state points. We now proceed to test the method in applications to multi-component systems with more complex, rugged, and challenging-to-sample free energy surfaces.

\subsection{Association/dissociation of a NaCl ion pair in water}

We now apply PINES to discover CVs to drive the association/dissociation of a NaCl ion pair in water at 300 K. It is well known that solvent degrees of freedom play an important role in mediating this process via many-body collective rearrangements of solvation shell water molecules \cite{jung2019artificial,mullen2014transmission,kellermeier2016entropy,geissler1999kinetic,wang2022influence}. We hypothesize that PINES should be able to identify permutationally-invariant ion and solvent-inclusive CVs associated with the indistinguishable solvent molecules and use these to drive sampling of the solvated ion pair dynamics.

\begin{figure}[ht!]
\centering
  \includegraphics[width=0.86\textwidth]{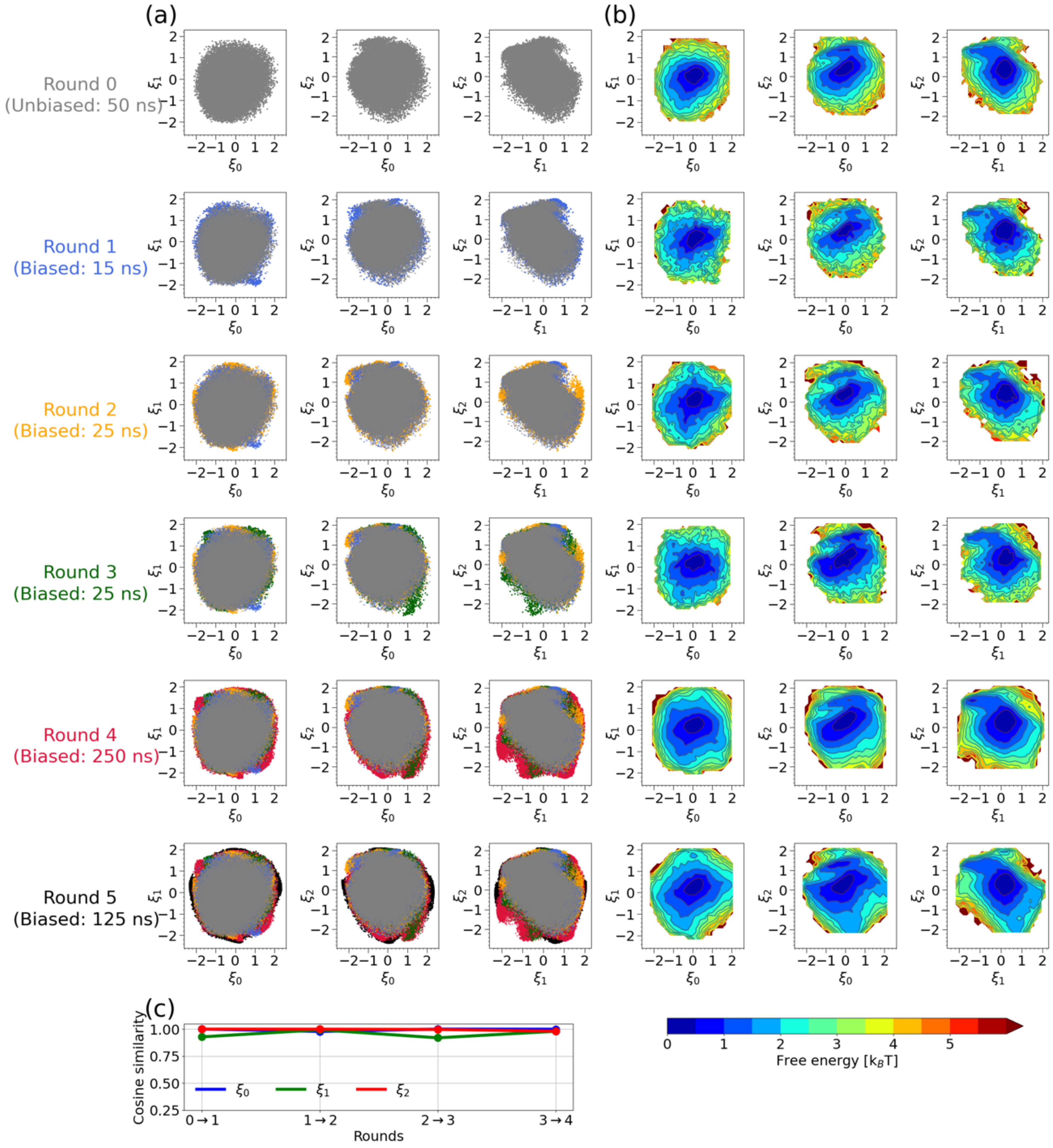}
  \caption{Application of PINES to a NaCl ion pair in water at 300 K. (a) Each column shows a 2D embedding of the round-by-round accrued sampling projected into the converged CVs $\xi_{0}$, $\xi_{1}$, and $\xi_{2}$ at the terminal Round 5. The unbiased simulation quite thoroughly samples the thermally accessible phase space, yet each biased round advances the frontier into unsampled regions on the periphery of the explored region. Rounds 1-3 show broader exploration of the CV space in half or less of the simulation time compared to the unbiased run and indicate that the CVs are accelerating sampling of novel states. Rounds 4-5, despite being much longer than the previous biased runs, provide diminishing returns in the exploration of new areas of CV space. (b) Corresponding free energy surfaces (FES) computed by reweighting of the biased simulation data collected in the current round of sampling. (c) Cosine similarity between corresponding CVs in successive rounds. The high cosine similarity values between all neighboring rounds indicate that the CVs discovered in the initial unbiased round are not substantially altered by learning over the biased data and the PINES protocol rapidly converges.}
  \label{Fig6}
\end{figure}

We identified a $d$=3 dimensional latent space from an initial 50 ns unbiased simulation trajectory and confirm that this dimensionality remained unchanged in each of the five successive rounds of PINES. We illustrate in Fig.~\ref{Fig6}a the expansion of the sampled phase space under each PINES iteration. The initial unbiased simulation provides good initial sampling of the central portion of the latent space (gray) but each successive biased round (different colors) provide additional sampling of the periphery. In particular, Rounds 1-3 (blue, yellow, green) show increased exploration of the CV space in half or less of the simulation time compared to the unbiased run and indicate that the CVs are accelerating sampling of novel states. Rounds 4-5 (red, black), despite being longer than the previous runs, offer diminishing returns in additional configurational sampling. (We recall that the run length in each round is dictated by convergence criteria of the PBMetaD simulations (cf.\ Sec.~\ref{sec:PBMetaD}).) 

In Fig.~\ref{Fig6}b, we present the corresponding reweighted FES for each round. We observe visually apparent expansion of the FES into high-free energy peripheral regions and changes in the free energy distributions through Rounds 0-4. The FES between Rounds 4-5, however, are very similar in terms of the explored regions and free energy contours. As we discuss below, the core of the latent space that was readily sampled by the unbiased simulation corresponds to dissociated configurations with average levels of hydration. The peripheral states that require biased simulations for adequate exploration are associated with higher free energy states corresponding to metastable configurations such as the solvent-shared pair.

We present in Fig.~\ref{Fig6}c the cosine similarity between CVs in successive PINES rounds. In this case we find that he CVs almost immediately stabilize, with those learned from the unbiased simulation data changing relatively little over the course of the iterative CV discovery and enhanced sampling protocol and always remaining above $\rho$ = 0.9. Taken together with the observed trends in the latent space and FES, we surmise that the initial CVs provide a good global parameterization of the system configurational space, but require approximately three rounds of enhanced sampling and minor CV refinement to attain convergence in sampling of the thermally accessible phase space.

\begin{figure}[ht!]
\centering
  \includegraphics[width=1\textwidth]{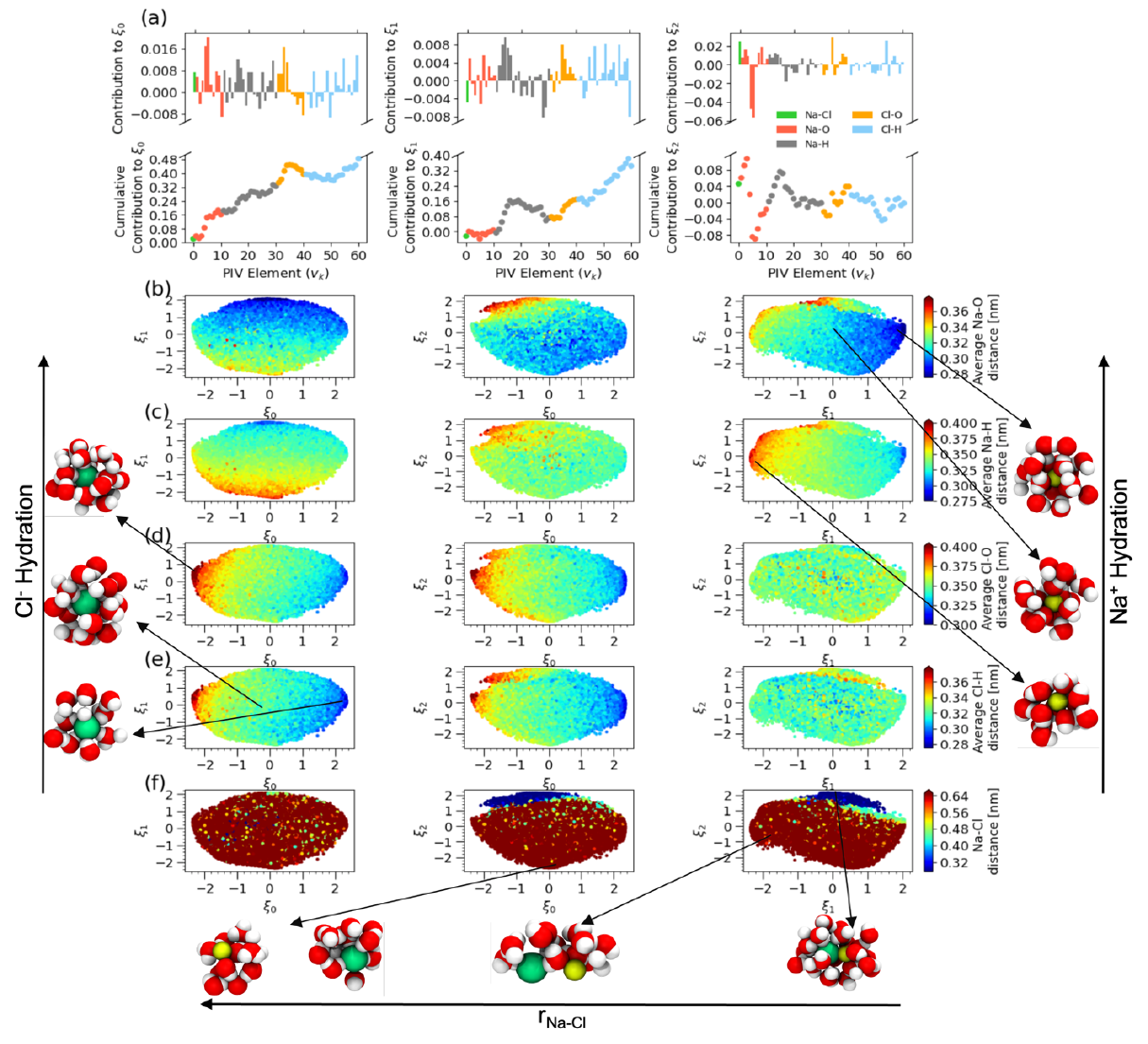}
   \caption{Bottom-up and top-down analysis of terminal learned CVs for the NaCl ion pair in water. (a) Split plots show the signed contribution from each PIV element to each CV $\{ \xi_0, \xi_1, \xi_2 \}$ as a bar chart on top and as a cumulative sum of the signed contributions on the bottom. The magnitude of contributions on the bar chart emphasize the importance of individual PIV elements while the slopes in different regions of the scatter plot illuminate the concerted effect of atom groups on each CV. (b-f) Projections of the terminal Round 5 sampling data into the terminal CVs and colored by physically interpretable candidate variables. $\xi_{0}$ is well-correlated with the average Cl-O and Cl-H distance, $\xi_{1}$ with the average Na-O and Na-H distance, and $\xi_2$ with the Na-Cl distance. Three distinct color bands are apparent in panel (f) corresponding to the contact pair (blue), the solvent-shared (green), and disassociated (red) states (cf.~Fig.~\ref{Fig8}c).}
   \label{Fig7}
\end{figure}

In Fig.~\ref{Fig7}a, we expose the integrated weights linking each of the 61 PIV elements to each of the three CVs. Inspection of the individual (upper) and cumulative (lower) weights indicates that particular interactions appear to contribute more significantly to each of the three CVs. In particular, PIV elements corresponding to Na-O and Cl-O distances make large magnitude contributions to $\xi_0$, whereas Na-H and Cl-H distances seemingly contribute more strongly to $\xi_1$, and Na-O makes the largest contribution to $\xi_2$.  Interestingly, the Na-Cl interaction is not a top contributor to any CVs, suggesting that the information about the Na-Cl state may be delocalized among the information on solvation shell water distances. 

We further explore these trends in Fig.~\ref{Fig7}b-f where we color the latent space by candidate physical variables. These plots are illuminating in showing that $\xi_{0}$ is well-correlated with the average Cl-O and Cl-H distance, $\xi_{1}$ with the average Na-O and Na-H distance, and $\xi_2$ with the Na-Cl distance. This analysis offers a straightforward interpretation of the three learned CVs as corresponding to Cl ion hydration ($\xi_{0}$), Na ion hydration ($\xi_{1}$), and Na-Cl proximity ($\xi_{2}$). 

\begin{figure}[ht!]
\centering
  \includegraphics[width=1\textwidth]{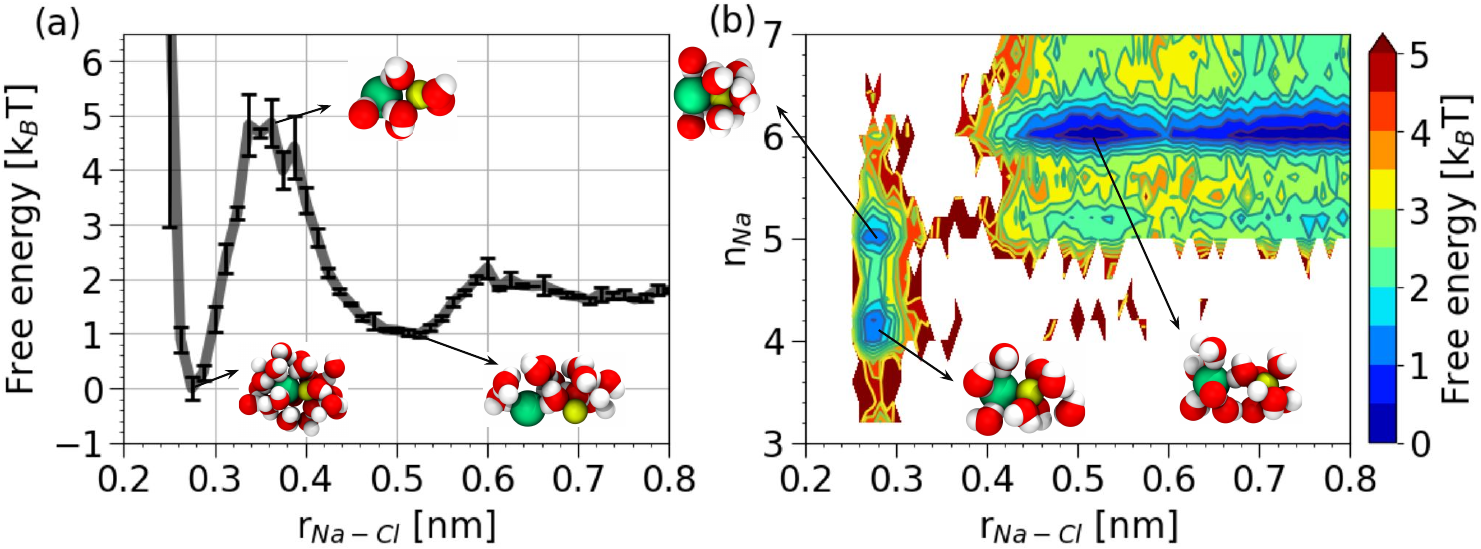}
  \caption{Reweighted free energy surfaces from the terminal round of enhanced sampling for the NaCl ion pair in water projected into physically interpretable order parameters. (a) A 1D FES projected into the Na-Cl distance. The landscape clearly illustrates the contact pair ($r_\text{Na-Cl}$ = 0.26 nm), solvent-shared ($r_\text{Na-Cl}$ = 0.51 nm), and disassociated ($r_\text{Na-Cl}$ $\gtrsim$ 0.6 nm) states. The free energy barrier lying between the contact pair and solvent-shared states is (4.9$\pm$0.1) $k_B T$ as measured from the contact pair, and involves a collective rearrangement of solvent shell water molecules to accommodate the separating ion pair \cite{geissler1999kinetic,Dellago:Committor}. Uncertainties in the free energy correspond to standard errors estimated by five-fold block averaging. (b) A 2D FES projected into the Na-Cl distance and the solvent coordination number of the Na ion $n_\text{Na}$. The contact pair comprises multiple solvation states of the Na ion. Transitions to the solvent separated pair appear to proceed most readily through the $n_\text{Na}$ = 5 contact pair state.}  
  \label{Fig8}
\end{figure}

We present in Fig.~\ref{Fig8} the reweighted FES collected from our final round of PINES enhanced sampling projected into physical order parameters and annotated with representative molecular configurations. A 1D FES constructed in the Na-Cl distance clearly shows the contact pair ($r_\text{Na-Cl}$ = 0.26 nm), solvent-shared ($r_\text{Na-Cl}$ = 0.51 nm), and disassociated ($r_\text{Na-Cl}$ $\gtrsim$ 0.6 nm) states (Fig.~\ref{Fig8}a). The free energy barrier between the contact pair and solvent-shared state lies at $r_\text{Na-Cl}$ = 0.36 nm with a height relative to the contact pair of (4.9$\pm$0.1) $k_B T$. This stands in good agreement with reported literature values of 4.8-5.3 $k_B T$ and provides good support for the accurate sampling of the configurational space and FES generated by PINES \cite{ssagessoftwaresuite,TimkoNaCldissociation}. We present in Fig.~\ref{Fig8}b a 2D FES constructed in the Na-Cl distance and the solvent coordination number of the Na ion $n_\text{Na}$. This projection is illuminating in showing a splitting of the contact pair at $r_\text{Na-Cl}$ = 0.26 nm into two basins with coordination numbers of $n_\text{Na}$ = 4 and $n_\text{Na}$ = 5. The FES suggests that the transition to the solvent-shared state at $r_\text{Na-Cl}$ = 0.51 nm appears to proceed from the $n_\text{Na}$ = 5 state and then drops into the lower free energy $n_\text{Na}$ = 6 state after separation has completed, although dynamical analyses such as transition path sampling would be required to test this conjecture.

To assess the transferability of the converged CVs learned by PINES for the NaCl ion pair association/dissociation process, we conducted enhanced sampling calculations in these CVs for a hypothetical divalent ion system generated by doubling the charge on each ion (Fig.~S5). Whereas an unbiased simulation of the divalent ion pair is completely localized near the initial state corresponding to the contact pair because of the strong electrostatic attractions (Fig.~S5a), PBMetaD enhanced sampling simulations in the monovalent CVs are capable of surmounting a $\sim$45 $k_B T$ free energy barrier to drive sampling of the dissociated state (Fig.~S5b). 
This analysis indicates the transferability of CVs learned over one system applied to another in which it is more challenging to drive sampling.

In sum, this application to a NaCl ion pair in water demonstrates the capacity of PINES to identify permutationally-invariant CVs incorporating information on the ion pair and local solvation environment and use these variables to efficiently drive enhanced sampling of ion pair association/dissociation. This illustrates the capacity of the method to learn permutationally-adapted CVs in a solvated system of chemically indistinguishable water molecules. The unbiased simulations used to seed the PINES campaign provided a relatively comprehensive sampling of the phase space and the learned CVs remained relatively unchanged over the course of the iterative search. In the next application, we demonstrate a case where the PINES enhanced sampling is instrumental in good sampling of phase space.

\subsection{ \texorpdfstring{Hydrophobic collapse of a \textit{n}-pentatetracontane (\ce{C_{45}H_{92}}) polymer chain in water}{Hydrophobic collapse of a n-pentatetracontane (C45H92) polymer chain in water}}

In our final application, we consider hydrophobic collapse of an \textit{n}-pentatetracontane (\ce{C_{45}H_{92}}) polymer chain in water to demonstrate an application of PINES to a solvated polymer system. The dynamical collapse of hydrophobic polymers in water is known to be intimately related to drying transitions and the formation of water cavities  \cite{ten2002drying,athawale2007effects,ferguson2010sys,Patel2021PolymerSolvationINDUS}. Solvent-centric CVs are therefore expected to play an important role in the collapse/extension transitions of the \ce{C_{45}H_{92}} polymer chain and, for such a long chain, it can also be challenging to comprehensively explore the configurational phase space using unbiased simulations alone. This system is therefore anticipated to be a good test of the PINES methodology.

\begin{figure}[ht!]
\centering
  \includegraphics[width=0.95\textwidth]{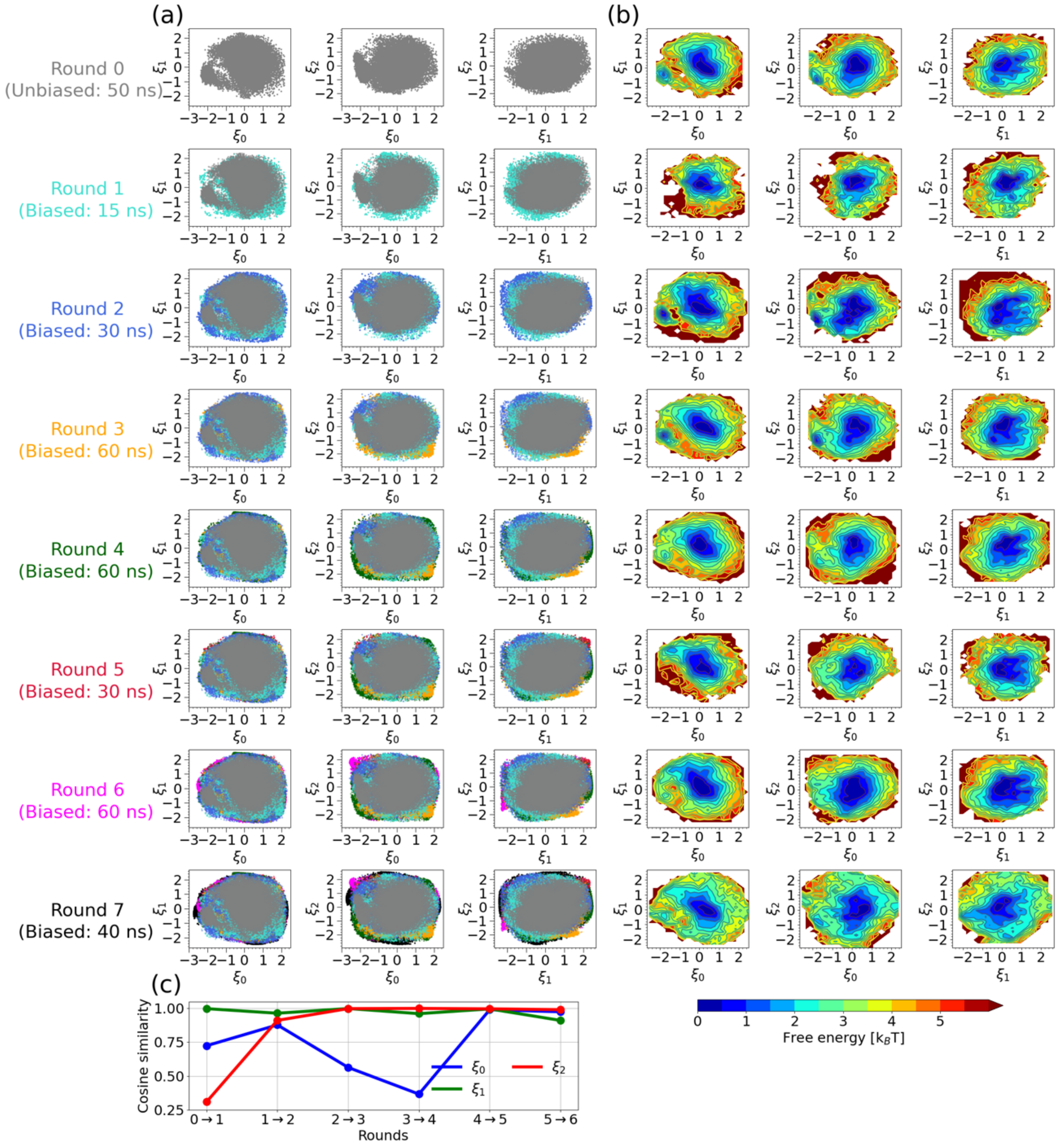}
  \caption{Application of PINES to an \textit{n}-pentatetracontane (\ce{C_{45}H_{92}}) chain in water at 300 K. (a) Each column shows a 2D embedding of the round-by-round accrued sampling projected into the converged CVs $\xi_{0}$, $\xi_{1}$, and $\xi_{2}$ at the terminal Round 7. Rounds 1-4 show significantly increased exploration of novel states compared to the unbiased run and indicate that the CVs are accelerating sampling of unexplored regions of phase space. In Rounds 5-6 the sampling of the CV space begins to saturate and suggests that the CVs have converged. (b) Corresponding free energy surfaces (FES) computed by reweighting of the biased simulation data collected in the current round of sampling. (c) Cosine similarity between corresponding CVs in successive rounds. The CVs change significantly round-to-round over Rounds 1-4 before converging in Rounds 5-7.}
  \label{Fig9}
\end{figure}

Once again, we identified a $d$=3 dimensional latent space from and initial 50 ns unbiased trajectory that remained unchanged over the course of the seven subsequent rounds of PINES. We present in Fig.~\ref{Fig9}a the expansion of the sampled latent space over the PINES iterations. We see significant expansion into new, unexplored regions by the enhanced sampling calculations in Rounds 1-4, before sampling approximately converges and achieves only minor additional exploration in Rounds 5-7. We draw particular attention to the interior ``hole'' in the latent space apparent at ($\xi_0$=-1, $\xi_1$=0) in Round 0 (gray) and which is filled in by the enhanced sampling trajectories in Round 1 (cyan) and Round 2 (blue). As we discuss below, this hole corresponds to a high-free energy transition region between collapsed and extended states of the chain that is only adequately sampled under PBMetaD biasing in the learned CVs. Fig.~\ref{Fig9}b presents analogous FES plots for each PINES round that expose the density distribution of points over the latent space and convergence of the landscapes over the PINES rounds. Fig.~\ref{Fig9}c presents the cosine similarity between CVs in successive PINES rounds. We observe significant changes in the learned CVs over Rounds 1-4 as they are exposed to additional configurational data supplied by the enhanced sampling calculations. The CVs stabilize in Rounds 5 and 6 with all CVs maintaining cosine similarity values of $\rho$ $>$ 0.9.

\begin{figure}[ht!]
\centering
  \includegraphics[width=1\textwidth]{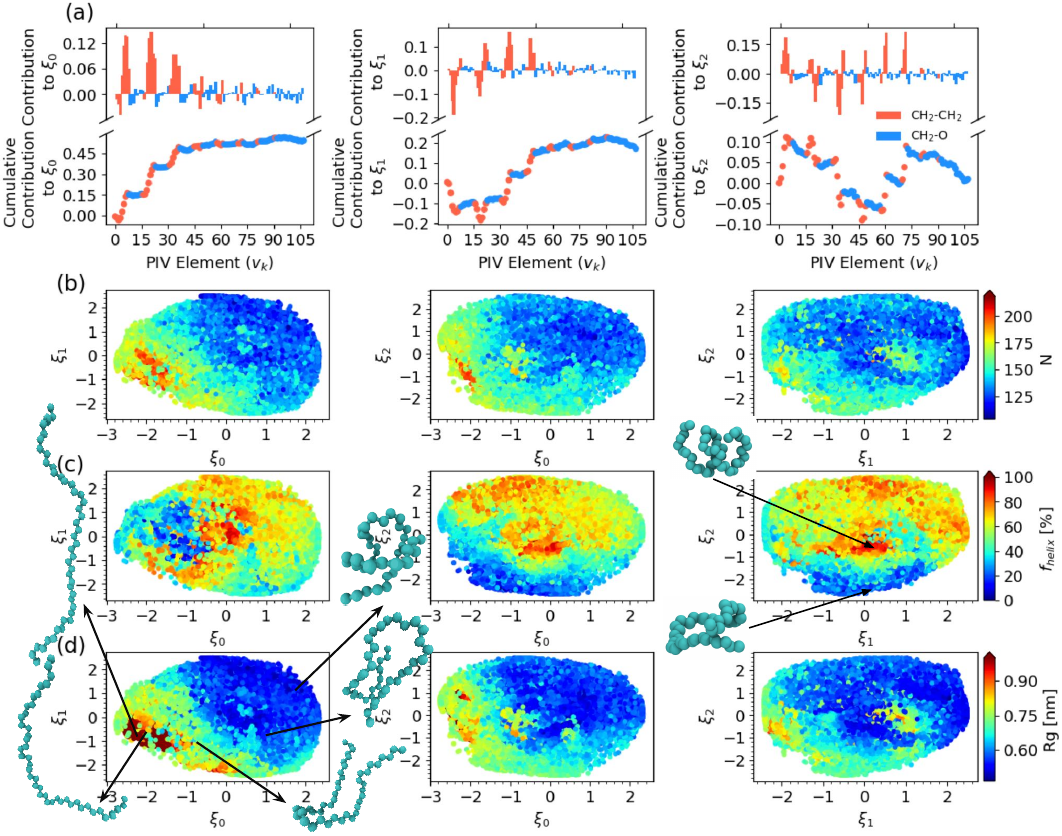}
  \caption{Bottom-up and top-down analysis of terminal learned CVs for the \textit{n}-pentatetracontane chain in water. (a) Split plots show the signed contribution from each PIV element to each CV $\{ \xi_0, \xi_1, \xi_2 \}$ as a bar chart on top and as a cumulative sum of the signed contributions on the bottom. The magnitude of contributions on the bar chart emphasize the importance of individual PIV elements while the slopes in different regions of the scatter plot illuminate the concerted effect of atom groups on each CV. (b-f) Projections of the terminal Round 7 sampling data into the terminal CVs and colored by physically interpretable candidate variables. $\xi_{0}$ and $\xi_1$ are correlated with the chain radius of gyration $R_g$ the number of waters $N$ in the first solvation shell of the chain. We highlight representative configurations on the $\xi_0$-$\xi_1$ surface corresponding to the extended (red), collapsed (blue), and partially collapsed (green) structures. $\xi_2$ is correlated with the helical pitch $f_\text{helix}$. We highlight representative configurations on the $\xi_1$-$\xi_2$ surface corresponding to planar rings (blue) and corkscrews (red).}
  \label{Fig10}
\end{figure}

In Fig.~\ref{Fig10}, we gain insight into the learned CVs by presenting the network weights linking each of the 108 PIV elements to the three CVs and coloring the latent space embeddings by candidate physical variables. Interpreting the CVs for more complex systems can be challenging, but we find $\xi_{0}$ and $\xi_1$ to possess significant correlation with both the chain radius of gyration $R_g$ and the number of waters $N$ in the first solvation shell of the chain, defined by counting the number of water molecule O atoms within 0.6 nm of any united atom \ce{CH2} or \ce{CH3} bead of the alkane chain. $\xi_2$ is correlated, albeit non-monotonically, with the fraction of helical pitch $f_\text{helix}$. This observable, defined as $f_\text{helix}$ = $r_{CH3 - CH43}$ - ($r_{CH3 - CH23}$ + $r_{CH13 - CH33}$ + $r_{CH23 - CH43}$) and min-max normalized, measures deviations from a common helical chain structure with a 20 united atom turn, where $r_{CH3 - CH43}$ is the distance between the 3\textsuperscript{rd} and 43\textsuperscript{rd} united atoms in the chain and ($r_{CH3 - CH23}$ + $r_{CH13 - CH33}$ + $r_{CH23 - CH43}$) are the distances between 20 united atom separations in the chain. Large values of $f_\text{helix}$ correspond to an extended chain with regular turns of 20 united atoms while low values correspond to wider helices and ring-like structures. It also appears that solute-solvent PIV elements are most strongly represented within $\xi_2$. Taken together, the CVs appear to contain signatures of the compactness, global hydration, and helicity of the polymer chain within the water solvent.

\begin{figure}[ht!]
\centering
  \includegraphics[width=1\textwidth]{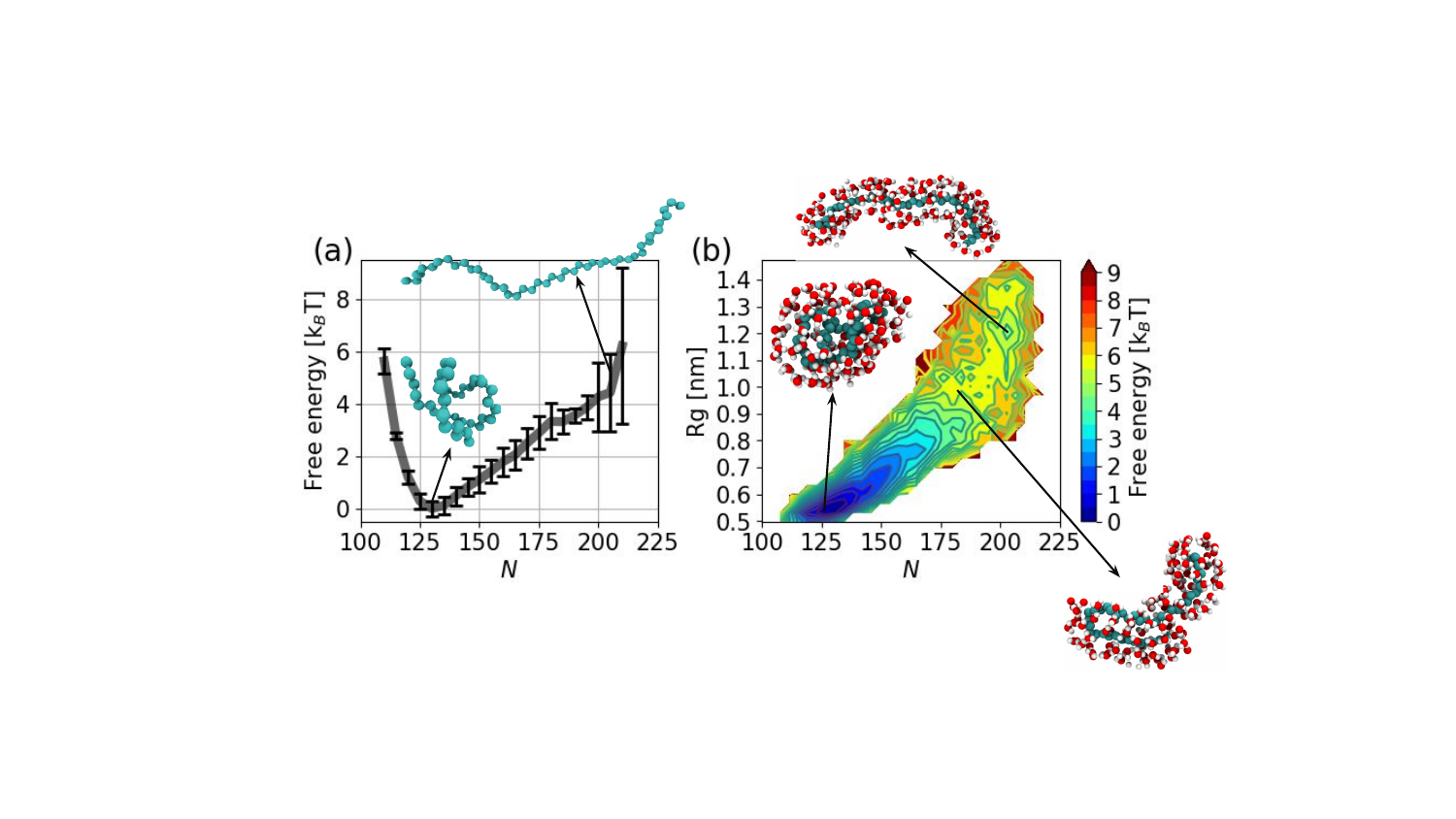}
  \caption{Reweighted free energy surfaces from the terminal round of enhanced sampling for the \textit{n}-pentatetracontane chain in water. (a) A 1D FES projected into the number of waters $N$ in the first solvation shell of the chain. A collapsed state with $N$$\approx$130 is more stable than the fully extended state with $N$$\approx$200 by approximately 4 $k_B T$. Uncertainties in the free energy correspond to standard errors estimated by five-fold block averaging. (b) A 2D FES projected into the number of waters $N$ in the first solvation shell of the chain and the radius of gyration $R_g$. The global free energy minimum at ($N$$\approx$130, $R_{g}$$\approx$0.57 nm) contains the hydrophobically collapsed chain.}
  \label{Fig11}
\end{figure}



In Fig.~\ref{Fig11} we show our converged and reweighted FES from the terminal round of PINES projected into physical order parameters. A 1D FES projected into the number of waters $N$ in the first solvation shell of the chain exposes the collapsed state of the chain to be more stable than the extended by approximately (4$\pm$1) $k_B T$, in good agreement with the value of 3.3 $k_B T$ reported by Dhabal \textit{et al.} \cite{Patel2021PolymerSolvationINDUS} (Fig.~\ref{Fig11}a). A 2D FES constructed in $N$ and $R_g$ clearly exposes the relationship between hydration and chain extent, with the collapsed state of the chain better shielding the hydrophobic beads from solvent and occupying the global free energy minimum at this thermodynamic state point (Fig.~\ref{Fig11}b).

\section{\label{sec:conclusions}Conclusions}

In this work we have introduced Permutationally Invariant Network for Enhanced Sampling (PINES) as an approach for data-driven discovery and enhanced sampling in translationally, rotationally, and permutationally invariant collective variables for molecular systems. The approach essentially integrates Permutation Invariant Vector (PIV) featurizations with the Molecular Enhanced Sampling with Autoencoders (MESA) approach to perform interleaved rounds of symmetry-adapted collective variables and parallel bias metadynamics enhanced sampling to perform comprehensive sampling of the thermally accessible phase space and obtain converged thermodynamic averages. Convergence of the iterative sampling process is assessed by stabilization of the learned collective variables and saturation of the explored configurational phase space. PINES is well-suited to data-driven collective variable discovery and enhanced sampling in challenging to sample systems of indistinguishable particles frequently encountered in self-assembling and solvated systems. We have demonstrated PINES in applications to the assembly/disassembly of a 13-particle Ar cluster, association/dissociation of a NaCl ion pair in water, and hydrophobic collapse of a \textit{n}-pentatetracontane (\ce{C_{45}H_{92}}) chain in water. In all instances, PINES efficiently discovers symmetry-adapted multi-body collective variables that are used to efficiently drive sampling of configurational space. We anticipate that PINES may be of value in applications to diverse molecular systems comprising indistinguishable particles and we have made the approach freely available as a new user-friendly module at \url{https://github.com/Ferg-Lab/pines.git} that can be patched with PLUMED2 enhanced sampling libraries and will be submitted for incorporation into a future release of PLUMED2 \cite{PLUMED2}.


In future work, we would like to apply PINES to engage more complex processes such as protein-folding, DNA self-assembly, and ligand-binding. Additionally, we envision a number of future technical improvements and modifications to PINES. One of the key advantages of PINES is its flexibility through modularity. Any generic CV-based biasing method is immediately compatible with PINES, but with small modifications to the open-source code the featurization choice and the network architecture may also be adapted to employ alternative CV discovery techniques such as, for example, SRVs to discover slow, as opposed to high-variance, CVs. We also note that a PyTorch PLUMED2 module \cite{Bonati:20:DatadrivenCVJPCL,Bonati:PlumedTorch:Web} enables extension of the PINES approach to use arbitrarily complicated ANN architectures within the PINES pipeline. In addition, we are also implementing PINES in the flexible GPU accelerated enhanced sampling library PySAGES \cite{rico2023pysages}. A principal use case of PINES is its application to solvated systems. The current implementation of PINES employs only solute-solvent distances within the PIV and does not consider the solvent-solvent interaction block. While this has the benefit of reducing the size of the PIV, the inclusion of such interactions would provide additional information on the structure of the near-field solvent that may be important for particular systems such as proteins with highly structured solvation shells. Lastly, we note that the current method makes no use of the decoder portion of the network other than its necessity in training the autoencoder network. The decoder, however, could possibly be utilized to generate novel conformations by sampling from sparse regions in the latent space to seed enhanced sampling calculations and accelerate convergence of the PINES protocol.

\section*{Conflict of Interest Disclosure}

A.L.F.\ is a co-founder and consultant of Evozyne, Inc.\ and a co-author of US Patent Applications 16/887,710 and 17/642,582, US Provisional Patent Applications 62/853,919, 62/900,420, 63/314,898, 63/479,378, and 63/521,617, and International Patent Applications PCT/US2020/035206 and PCT/US2020/050466.

\begin{acknowledgement}

Work by N.S.M.H.\ was supported by the National Science Foundation under Grant No.\ DMR-1841807. Work by S.D.\ was supported by MICCoM (Midwest Center for Computational Materials), as part of the Computational Materials Science Program funded by the U.S. Department of Energy, Office of Science, Basic Energy Sciences, Materials Sciences and Engineering Division, through Argonne National Laboratory, under Contract No.\ DE-AC02-06CH11357. This work was completed in part with resources provided by the University of Chicago Research Computing Center. We gratefully acknowledge computing time on the University of Chicago high-performance GPU-based cyberinfrastructure supported by the National Science Foundation under Grant No.\ DMR-1828629.

\end{acknowledgement}

\begin{suppinfo}
Supplementary information comprising switching function parameterization process from pairwise RDFs, example of the FVE plots used to determine the appropriate dimensionality of the latent space, analysis of convergence and converged FES for 13-atom Argon system in an entropically dominated 5$\times$5$\times$5~nm$^{3}$ box, analysis of transferability of learned CVs for the 13 atom Argon system to out-of-training temperature regimes, and analysis of transferability of learned CVs from the NaCl system to a divalent ion pair by scaling the \ce{Na+} and \ce{Cl-} valency.
\end{suppinfo}


\clearpage
\newpage

\bibliography{ref}


\clearpage
\newpage

\begin{tocentry}
\centering
\includegraphics[width=2.8in]{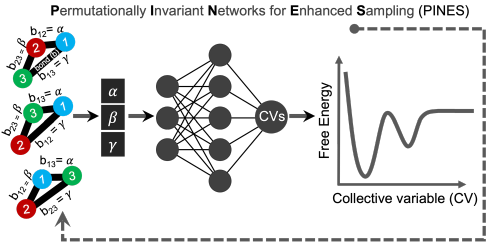}
\end{tocentry}

\end{document}